\documentclass[twocolumn,twocolappendix,trackchanges]{aastex7}
\usepackage[version=4]{mhchem}
\usepackage{amsmath}
\usepackage[utf8]{inputenc}
\shorttitle{Emission from Massive Protostellar Disks with Large Silicate Grains}
\shortauthors{Yamamuro et al.}

\graphicspath{{./}{figures/}}

\begin{document}

\title{The Impact of Silicate Grain Coagulation on Millimeter Emission from Massive Protostellar Disks}

\author[0000-0001-7530-1359]{Ryota Yamamuro}
\email{yamamuro.r.aa@m.titech.ac.jp}
\affiliation{Department of Earth and Planetary Sciences, Institute of Science Tokyo, Meguro, Tokyo, 152-8551, Japan}

\author[0000-0002-6907-0926]{Kei E. I. Tanaka}
\email{kei.tanaka@eps.sci.titech.ac.jp}
\affiliation{Department of Earth and Planetary Sciences, Institute of Science Tokyo, Meguro, Tokyo, 152-8551, Japan}

\author[0000-0002-1886-0880]{Satoshi Okuzumi}
\email{okuzumi@eps.sci.titech.ac.jp}
\affiliation{Department of Earth and Planetary Sciences, Institute of Science Tokyo, Meguro, Tokyo, 152-8551, Japan}

\begin{abstract}
Hot accretion disks around massive protostars provide a unique opportunity to study ice‐free silicate grains that cannot be investigated in protoplanetary disks.
We conduct a self‐consistent investigation into grain size evolution and its impact on (sub-)millimeter-wave emission from massive protostellar disks.
Our radiative transfer modeling accounts for dust self-scattering and includes vertical temperature gradients in the disk structure.
The results show that once silicate grains grow to sizes exceeding the observing wavelength, enhanced scattering dims the disk emission by $20$--$30\:\%$ relative to the blackbody emission expected at the disk surface temperature.
By comparing our model with Atacama Large Millimeter/submillimeter Array 1.14 mm observations of the disk around the massive protostar GGD27‑MM1,
we constrain the threshold velocity for collisional fragmentation of silicate grains to approximately $15{\rm\:m\:s^{-1}}$.
This fragmentation velocity is lower than the typical maximum collisional velocities in protoplanetary disks around low-mass stars, suggesting that collisional coagulation alone is insufficient for silicate dust to form rocky planetesimals in such environments.
Furthermore, our analysis identifies two potential scenarios to better reproduce the bright inner-disk emission of GGD27‑MM1.
One possibility is that the grain growth is limited to $160{\rm\:\mu m}$ by another growth barrier (e.g., collisional bouncing), reducing scattering dimming.
Alternatively, the stellar luminosity may be as much as five times higher than current estimates, compensating for the reduced brightness. 
Future multiwavelength observations, particularly at shorter submillimeter wavelengths, will be crucial to distinguish between these scenarios
and further constrain silicate grain coagulation processes in massive protostellar disks.
\end{abstract}


\section{Introduction} \label{sec:intro}
One of the key challenges in planet formation is understanding how kilometer-sized planetesimals emerge from submicron-sized dust grains.
While the fundamental processes of dust evolution through collisional coagulation have been intensively investigated, it has been suggested that collisional coagulation alone may be insufficient for grains to grow into planetesimals \citep{Birnstill2023}. 
A significant barrier against dust growth is collisional fragmentation, where dust grains are destroyed upon high-velocity impacts rather than {sticking} \citep{Ryan1991, Blum+1993}.
At sufficiently high collisional velocities, this fragmentation limits further grain growth, preventing grains from reaching planetesimal sizes.
Overcoming this barrier requires the gravitational collapse of dust overdensities, which may be formed by the streaming instability and (secular) gravitational instability \citep[e.g.,][]{Goldreich+1973, Youdin+2005,Youdin2011, Tominaga2018}.
Still, these instabilities necessitate some degree of grain growth for the grains to marginally decouple from gas motion and concentrate.
Thus, even if collisional growth alone cannot directly lead to planetesimal formation, it is essential to understand how far dust grains can grow through coagulation.

Both experimental and theoretical studies have investigated the sticking properties of dust grains.
Early studies suggested that $1~\mathrm{\mu m}$-sized silicate grains can stick at collision velocities up to $\sim 0.1$--$1~\mathrm{m\:s^{-1}}$ \citep{Chokshi+93, Dominik1997, Blum2000, Poppe+2000}.
However, more recent studies indicate that well-dried silicate grains are much stickier \citep{Kimura2015,Stein2019,Pillich2021}, possibly because adsorbed water vapor increases the intersurface distance between the grains in contact \citep{Nagaashi2021}.
Additionally, the stickiness of dust aggregates generally increases with the increasing size of the constituent grains \citep{Chokshi+93, Dominik1997}.
\citet{Kimura2015} predict that aggregates made of $0.1~\mathrm{\mu m}$ sized well-dried silicates can stick up to $50 ~\rm m~s^{-1}$. 

Given the varied conclusions from theory and observations, observational constraints from protoplanetary disks are essential to understanding the coagulation properties of silicate grains.
However, in protoplanetary disks, the region inside the snowline, where dust is expected to be rocky silicate, is considered to be only a few astronomical units in size \citep{Hayashi+81,Sasselov2000,Mori+21}.
Even with the capabilities of the Atacama Large Millimeter/submillimeter Array (ALMA), spatially resolving this region and obtaining information about the size and coagulation properties of silicate grains remains challenging.
Therefore, it is still uncertain how large silicate grains can grow within disks and what their coagulation properties are.

To address this challenge and constrain the coagulation properties of silicate grains from observations, some recent studies have focused on hotter disks with extensive rocky dust regions.
Disk temperatures can rise significantly due to accretion, as seen in FU Orionis-type outbursts around low-mass protostars \citep{Liu2021} and in Class 0/I protostellar disks heated by accretion \citep{Zamponi2024}.

Our study focuses on accretion disks around massive protostars with $\gtrsim 10\:M_\odot$.
Massive protostellar disks are heated by high stellar luminosities of $\gtrsim10^4\:L_\odot$ and high accretion rates of $\sim10^{-4}$--$10^{-3}\: M_\odot\:\mathrm{yr}^{-1}$. As a result, the snowlines extend to distances of $\sim100$--$1000\:\mathrm{au}$, placing the entire disk within the snow line.
Despite the distance of massive protostellar disks being $\gtrsim 1\:\mathrm{kpc}$, their rocky dust regions are extensive enough to be resolved in (sub-)millimeter wavelength observations.
In particular, the massive protostellar disk GGD27-MM1 has been observed with high-resolution polarization at $1.14\:\mathrm{mm}$ using ALMA \citep{Girart2018}.
These observations detected polarization parallel to the disk minor axis in the southwestern part of the disk, approximately $100\:\mathrm{au}$ from the central star.
This pattern is commonly interpreted as originating from self-scattering by grains with a radius of $\sim100\:\mathrm{\mu m}$ \citep{Kataoka2015}.
Additionally, several organic molecular lines, which are thought to be produced by ice sublimation, have been observed extending up to $\sim600\:\mathrm{au}$ \citep{Fernandez2023}.
Based on these observations, the presence of grown silicate grains is suggested in the disk.
These observations provide an opportunity to constrain the coagulation properties of silicate grains.

To explore this, we developed a theoretical model of a massive protostellar disk where the coagulation and fragmentation of silicate grains are treated self-consistently \citep{Yamamuro2023}.
While this model constrained the stickiness required to reproduce the observed silicate grain sizes,
it had not yet been tested for its ability to reproduce the overall $1.14\:\mathrm{mm}$ dust continuum emission of the GGD27-MM1 disk.
Significant variations in dust opacity due to grain-size growth require careful consideration in the continuum modeling,
as they strongly affect the emergent intensity at (sub-)millimeter wavelengths \citep[e.g.,][]{Birnstiel2018, Ueda2023}.
In contrast, \cite{Anez+2020} developed a model to reproduce the $1.14\:\mathrm{mm}$ emission of the GGD27-MM1 disk.
Their model determined the grain size and its spatial distribution by fitting the continuum emission,
but these parameters were not necessarily consistent with disk turbulence or dust coagulation and fragmentation.
Therefore, the present study aims to self-consistently constrain both the coagulation properties of silicate grains and the parameters reproducing the overall $1.14\:\mathrm{mm}$ continuum emission.

The structure of this paper is as follows.
In Section \ref{sec:Method}, we describe the disk model, dust growth framework, and the radiative transfer method.
Section \ref{sec:Result} presents the disk structure with grain evolution and its emergent intensities.
In Section \ref{sec:bestfit}, we constrain the physical parameters that best reproduce the GGD27-MM1 observations.
Section \ref{sec:Dis} explores the implications of our findings, including their relevance to rocky planetesimal formation.
Finally, Section \ref{sec:sum} provides a summary of the study.

\section{Model} \label{sec:Method}
We present our model of massive protostellar disks with dust evolution, along with radiative transfer to evaluate their (sub-)mm continuum emission.
In Section \ref{model:disk}, we describe a steady disk model with a supply of gas and dust from the surrounding envelope.
For dust grains, key processes such as collisional growth, fragmentation, radial drift, and vertical settling are incorporated.
In Section \ref{model:rad}, we introduce a new analytical solution for radiative transfer that assumes isotropic scattering, allowing us to calculate the emergent intensity while incorporating the vertical temperature gradient of the disk.
Additionally, in Section \ref{model:op}, we present a dust opacity model that consistently accounts for grain size evolution,
which is applied to both the disk temperature calculation and radiative transfer.

\subsection{Disk Model}
\label{model:disk}
The disk model used in this study is primarily based on that developed by \cite{Yamamuro2023}.
We provide a briefly overview the gas and dust disk model here; further details can be found in Section 2 of \cite{Yamamuro2023}.

We consider a steady, axisymmetric disk around the central massive protostar with the mass of $M$.
Gas and dust are supplied to the outer edge of the disk from the surrounding envelope
with the supply rates of $\dot{M}_{\rm gas}$ and $\dot{M}_{\rm dust}=f_{\rm dg,ISM}\dot{M}_{\rm gas}$, respectively.
We adopt the dust-to-gas mass ratio of $f_{\rm dg,ISM}=0.01$.
Starting from the outer edge, defined by the disk radius $r_{\rm disk}$, we compute the physical properties of the disk at each radial distance $r$,
including the gas and dust surface densities, $\Sigma_{\rm gas}$ and $\Sigma_{\rm dust}$,
the midplane temperature $T_{\rm mid}$, and the maximum grain size $a_{\rm max}$.

The gas surface density follows the equation of continuity of the steady disk:
\begin{equation}
\dot{M}_\mathrm{gas} = 3 \pi \frac{\alpha {c_s}^2}{ \Omega_\mathrm{K} }\Sigma_\mathrm{gas} = {\rm const.},
\end{equation}
where
$\alpha$ is the dimensionless viscosity parameter \citep{Shakura1973}, 
$c_s=\sqrt{k_\mathrm{B} T_{\rm mid} / m_\mathrm{gas}}$ is the sound speed,
and $\Omega_\mathrm{K}=\sqrt{G M /r^3}$ is the Keplerian angular speed, respectively.
Here, $G$, $k_\mathrm{B}$, and $m_\mathrm{gas}$ represent the gravitational constant, the Boltzmann constant, and the mean molecular mass, respectively.
We decompose the $\alpha$ parameter into two components:
\begin{equation}
    \alpha=\alpha_\mathrm{GI} + \alpha_\mathrm{floor} ,
\end{equation}
where $\alpha_\mathrm{GI}$ accounts for turbulence driven by gravitational instability (GI),
and $\alpha_\mathrm{floor}$ represents the baseline viscosity due to other mechanisms..
Following \cite{Zhu2010}, we model $\alpha_\mathrm{GI}$ as
\begin{equation}\label{alpha_GI}
    \alpha_\mathrm{GI}=\mathrm{exp}\left( -Q^{4} \right),
\end{equation}
where
\begin{equation}
    Q=\frac{c_s \Omega_\mathrm{K} }{\pi G {\Sigma_\mathrm{gas}}},
\end{equation}
is the Toomre parameter for GI \citep{Toomre1964}.
When $Q\lesssim1.4$, the disk becomes gravitationally unstable, causing an increase in $\alpha$ and enhanced angular momentum transport.
We take $\alpha_\mathrm{floor}=0.01$, following \cite{Tanaka2014}.

For the temperature calculation, we consider both stellar irradiation and accretion heating.
The disk midplane temperature is approximated as
\begin{equation}
    T_\mathrm{mid}=\left( T^4_\mathrm{irr} + T^4_\mathrm{acc,mid} \right)^{1/4},
\end{equation}
where $T_\mathrm{irr}$ and $T_{\rm acc,mid}$ are the temperatures when irradiation heating and accretion heating dominate, respectively.
The irradiation-dominated temperature $T_\mathrm{irr}$ is estimated as
\begin{equation}
    T_\mathrm{irr}=1.5\times10^2\left(\frac{L}{L_\odot}\right)^{2/7}
\left(\frac{M}{M_\odot}\right)^{-1/7}
\left(\frac{r}{\mathrm{au}}\right)^{-3/7} \:\mathrm{K},
\end{equation}
where $L$ is the stellar luminosity \citep{Chiang1974}.
The accretion-dominated temperature $T_\mathrm{acc, mid}$ is given by
\begin{equation}\label{eq:Tam}
    T_\mathrm{acc, mid} = \left(\frac{\sqrt{3}}{4}+\frac{3}{16}\tau_\mathrm{R, tot}\right)^{1/4}\left(\frac{3GM\dot{M}_\mathrm{gas}}{8\pi\sigma_\mathrm{SB}r^3}\right)^{1/4},
\end{equation}
where $\sigma_\mathrm{SB}$ is the Stefan–Boltzmann constant
and $\tau_\mathrm{R, tot}$ is the Rosseland-mean optical depth \citep{Hubney1990}.
The Rosseland-mean optical depth is calculated as
\begin{equation}
    \tau_\mathrm{R, tot}=\kappa_\mathrm{R} \Sigma_\mathrm{dust},
\end{equation}
where $\kappa_\mathrm{R}$ is the Rosseland-mean opacity per dust mass (the details of opacities are described in Section \ref{model:op}).

In regions where the temperature exceeds approximately $1800{\rm\:K}$, dust grains begin to sublimate.
This sublimation reduces the dust surface density, which in turn lowers the optical depth and weakens accretion heating, thereby preventing further sublimation. 
As a result, the temperature stabilizes around the sublimation threshold \citep[e.g.,][]{Tanaka2011}.
To account for this effect, we impose an upper limit of $T_{\rm sub}=1800{\rm\:K}$ for the midplane temperature.
When this threshold is reached, we reduce $\Sigma_{\rm dust}$ to a level consistent with $T_{\rm mid}=T_{\rm sub}$, effectively simulating the influence of dust sublimation on the disk structure.
This sublimation effect occurs only in the innermost regions of the disk and is unlikely to significantly affect the continuum emission observed at (sub-)millimeter wavelengths.

To model the dust disk with grain evolution, we employ the single-size approach of \cite{Sato2016}.
This method assumes that, at each orbit $r$, the dust surface density is dominated by the largest grains.
Therefore, we track only the largest grains with a radius $a_\mathrm{max}(r)$.
In steady state, the equation of continuity for dust gives
\begin{equation}
    \dot{M}_\mathrm{dust}=2 \pi r \Sigma_\mathrm{dust} |v_\mathrm{r, dust}| = {\rm const.},
\end{equation}
where $v_\mathrm{r, dust}$ is the radial velocity of the largest grains, which includes the effect of the radial drift. 
For grain size evolution, we consider both coagulation and fragmentation due to collisions:
\begin{equation}
   \frac{da_{\rm max}}{dr}
=\xi_\mathrm{frag}\frac{a_{\rm max}}{3t_\mathrm{coll} v_\mathrm{r,dust}} ,
\end{equation}
where $t_\mathrm{coll}$ is the mean collisional time for the largest grains.
In estimating $t_\mathrm{coll}$, we account for the effect of dust vertical settling.
The factor of $\xi_\mathrm{frag}$ denotes the mass-fraction change in a single collision, described by
\begin{equation}
\xi_\mathrm{frag}=    \mathrm{min} \left(1,-\frac{\mathrm{ln}(\Delta v/v_\mathrm{frag})}{\mathrm{ln}5}\right), 
\end{equation}
where $\Delta v$ is the mean collisional velocity,
and $v_\mathrm{frag}$ is the threshold velocity above which a net mass loss of the largest grains occurs \citep{Okuzumi2016}.
Hereafter, we simply refer to $v_\mathrm{frag}$ as the fragmentation velocity.
We adopt the maximum grain radius of $a_\mathrm{max}=0.1\:\mathrm{\mu m}$ at the outer edge of the disk as the boundary condition.

From the outer edge of the disk inward, we numerically solve the basic equations presented above, evaluating both the gas disk structure and the dust evolution self-consistently.
In this study, we stop the calculations at an inner radius of $10{\rm\:au}$.

\subsection{Radiative Transfer for Disk Emission}
\label{model:rad}
Based on the disk model described above, we perform radiative transfer calculations of dust thermal emission to investigate how these disks would be observed at (sub-)mm wavelengths.
Given that the disk is geometrically thin, we approximate it as a vertically one-dimensional plane-parallel slab at each radial position.
To accurately evaluate the emission, we need to take into account the vertical thermal structure of the disk.
When accretion heating dominates, the temperature profile is given by \citep{Hubney1990}
\begin{equation}\label{eq:Taz}
     T_\mathrm{acc}(\tau_\mathrm{R})= 
     \left[
     \frac{\sqrt{3}}{4}+\frac{3 \tau_\mathrm{R}( \tau_\mathrm{R, tot}-\tau_\mathrm{R}) }{4\tau_\mathrm{R, tot}}
     \right]^{1/4}
     \left(\frac{3GM\dot{M}_\mathrm{gas}}{8\pi\sigma_\mathrm{SB}r^3}\right)^{1/4},
\end{equation}
where $\tau_\mathrm{R}= \int\kappa_{\rm R}d\Sigma_{\rm dust}$ is the vertical optical depth, used as the vertical coordinate.
Here, $\tau_\mathrm{R}=0$ corresponds to the disk surface facing away from the observer,
and $\tau_\mathrm{R}=\tau_\mathrm{R,\rm tot}$ denotes the observer-facing surface.
This formulation aligns with the accretion-dominated midplane temperature $T_{\rm acc,mid}$,
given $\tau_\mathrm{R}=\tau_\mathrm{R,\rm tot}/2$ at the midplane (see Equation \eqref{eq:Tam}).
Since irradiation heating does not produce a vertical profile, the combined temperature profile is obtained as
\begin{equation}
    T(\tau_{\rm R})=\left({T^4_\mathrm{irr}}+{T^4_\mathrm{acc}}(\tau_{\rm R})\right)^{1/4}.
\end{equation}
At the disk surfaces, $\tau_{\rm R}=0$ and $\tau_{\rm R,tot}$, the temperature is primarily determined by irradiation heating,
i.e., $T \approx T_{\rm irr}$.

We consider observing the disk at an inclination angle $i$ and frequency $\nu$ (or wavelength $\lambda=c/\nu$).
Here, we account for the effect of dust self-scattering,
which is known to be important for disk emission when grain sizes become comparable to or larger than the observational wavelength,
i.e., $a_\mathrm{max}\gtrsim \lambda/(2\pi )$ \citep[e.g.,][]{Bohren1983, Ueda2020}.
The radiative transfer equation along the ray to the observer is
\begin{equation}
    \mu \frac{d I_\nu(\tau_\nu)}{d \tau_\nu} = -I_\nu(\tau_\nu)
    + \epsilon^\mathrm{eff}_\mathrm{\nu}B_\nu(T(\tau_\nu))
    + \left( 1-\epsilon^\mathrm{eff}_\mathrm{\nu} \right)J_\nu(\tau_\nu),
\end{equation}
where $\mu = \cos\:i$ is the cosine of the inclination angle,
$\tau_{\nu}$ is the vertical optical depth,
$I_\nu$ is the intensity, $B_\nu(T)$ is the Planck function at a temperature $T$,
$J_\nu=\int I_\nu d\Omega/(4\pi)$ is the mean intensity {(with $d\Omega$ denoting the differential solid angle)},
and $\epsilon^\mathrm{eff}_\mathrm{\nu}$ is the absorption efficiency.
The optical depth $\tau_\nu$ is used as the vertical coordinate and proportional to the Rosseland-mean one:
\begin{equation}
    \tau_\nu=\frac{\kappa_{\nu,\rm abs}+\kappa^{\rm eff}_{\nu,\rm sca}}{\kappa_{\rm R}}\tau_{\rm R}.
\end{equation}
$\kappa_{\nu,\rm abs}$ and $\kappa^{\mathrm{eff}}_{\nu,\rm sca}$ are the absorption and scattering opacities, respectively.
The absorption efficiency is given by
\begin{equation}
    \epsilon^\mathrm{eff}_\mathrm{\nu}=\frac{\kappa_{\nu,\rm abs}}{\kappa_{\nu,\rm abs}+\kappa^{\rm eff}_{\nu,\rm sca}}.
\end{equation}
The superscript ``eff'' denotes an effective value for isotropic scattering (for more detail, see Section \ref{model:op}).

Incorporating scattering into radiative transfer calculations is computationally expensive, making it challenging to conduct broad parameter surveys or efficiently fit observational data.
For vertically isothermal disks,
previous studies on protoplanetary disks have used the analytical formula of the mean intensity $J_\nu$ derived by \cite{Miyake1983} to explore the effects of scattering \citep[e.g.,][]{Dalessio1998, Sierra2019, Zhu2019}.
Some studies have applied this isothermal solution to non-isothermal disks \citep[e.g.,][]{Dalessio2001,Sierra2020}.
However, this approach is inaccurate for vertically non-isothermal disks as the source function should then account for the entire vertical thermal profile.

To overcome the aforementioned limitation, we have derived a new analytical formula for the source function that handles vertically non-isothermal disks. 
Our method extends the approach of \cite{Miyake1983}, who applied the Eddington and two-stream approximations to derive the source function for isothermal disks,
by incorporating vertical temperature profiles to account for non-isothermal structures.
The detailed derivation of this analytical solution, along with a discussion of its accuracy, will be presented in a forthcoming paper (Tanaka et al., in prep.).
Here, we provide only the final form of the emerging intensity:
\begin{equation}
\begin{split}
&I_\nu(\tau_{\nu, \mathrm{tot}}) = \frac{\epsilon^\mathrm{eff}_\nu}{\mu} \frac{1 - 3\mu^2}{1 - 3\epsilon^\mathrm{eff}_\nu \mu^2} 
    \int^{\tau_{\nu, \mathrm{tot}}}_0 B_\nu (\tau_\nu) e^{-\tau_\nu/\mu} d\tau_\nu \\
    &+ \frac{\sqrt{3}\epsilon^\mathrm{eff}_\nu (1 - \epsilon^\mathrm{eff}_\nu)}{1 - 3\epsilon^\mathrm{eff}_\nu \mu^2}
 \frac{(1 + \sqrt{3}\mu) - (1 - \sqrt{3}\mu) e^{-\tau_{\nu, \mathrm{tot}}/\mu}}{(1 + \sqrt{\epsilon^\mathrm{eff}_\nu}) + (1 - \sqrt{\epsilon^\mathrm{eff}_\nu}) e^{-\sqrt{3\epsilon^\mathrm{eff}_\nu} \tau_{\nu, \mathrm{tot}}}} \\
&  \times  \int^{\tau_{\nu, \mathrm{tot}}}_0 B_\nu (\tau_\nu) e^{-\sqrt{3\epsilon^\mathrm{eff}_\nu} \tau_\nu} d\tau_\nu,
\end{split}
\end{equation}
where $\tau_\mathrm{\nu, tot} =(\kappa_\mathrm{\nu, abs}+ \kappa^\mathrm{eff}_\mathrm{\nu, sca})\Sigma_\mathrm{dust}$ is the total vertical optical depth at the observing frequency.
With this new analytical solution, we can now explore emission from vertically non-isothermal disks,
making radiative transfer with scattering more efficient for parameter surveys and model comparisons with observations.
{ It can be shown that, under the assumption of $B_\nu = {\rm const.}$,
this solution reduces to the isothermal form given by \citet[][Equations 3 and 4]{Sierra2024}.
\footnote{When comparing with the expressions in \citet{Sierra2024},
note that our absorption efficiency $\epsilon^\mathrm{eff}_\nu$ corresponds to $\epsilon_\nu^2$ in their notation.}}

\subsection{Dust Opacity}
\label{model:op}
The optical properties of dust grains determine both the thermal structure and observational emission of the disk.
In this study, we utilize the optical properties of dust grains derived by {\tt Optool} \citep{Dominik2021}.
This public code is designed to generate opacity by considering the constituent materials, size distribution, and structure of dust grains as input parameters.
Here, we assume that grains are spherical and compact.
The grain size distribution is assumed to follow a power law with an exponent of $-3.5$ 
{from the classical theory of fragmentation cascades \citep{Dohnanyi1969, Birn2011}, }
and the minimum grain radius is set to $a_{\rm min}=0.1{\rm\:\mu m}$.
We adopt ``astronomical silicate" as the representative silicate grain composition, with an internal dust density of $3.3 \:\mathrm{g\:{cm}^{-3}}$ \citep{Draine2003}.
Using this tool, we obtain the absorption opacity $\kappa_\mathrm{\nu, abs}$ and the scattering opacity $\kappa_\mathrm{\nu, sca}$ per unit dust mass at various frequencies $\nu$ as functions of the maximum grain size $a_{\rm max}$.

We have assumed isotropic scattering in the radiative transfer discussed in Section \ref{model:rad}.
However, forward scattering becomes more pronounced when the grain size is large, i.e., $2\pi a_\mathrm{max} \gtrsim \lambda$.
To mitigate the limitations of this approximation, we replace the scattering opacity with the effective value for isotropic scattering:
\begin{equation}
    \kappa^\mathrm{eff}_\mathrm{\nu, sca}=(1-g_\nu)\kappa_\mathrm{\nu, sca},
\end{equation}
where $g_\mathrm{\nu}$ is the forward scattering parameter {for spherical grains, reflecting their tendency to scatter light in the forward direction, as derived from $Z_{11}$ element of the scattering matrix} \citep{Heyney1941}.

\begin{figure}[t]
\begin{center}
\includegraphics[width=\hsize]{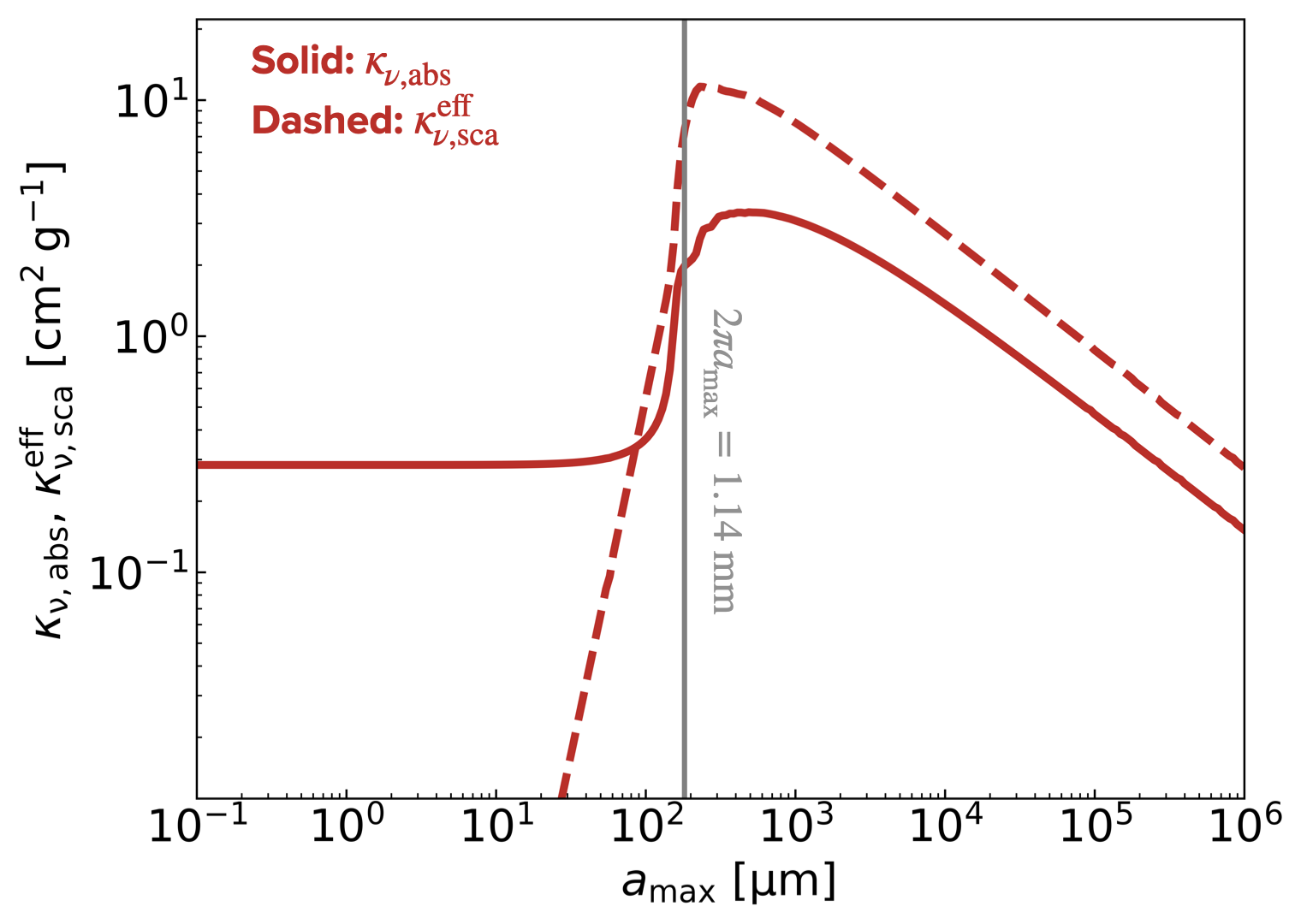}\
\caption{The absorption opacity $\kappa_\mathrm{\nu, abs}$ (solid line) and effective scattering opacity $\kappa^\mathrm{eff}_\mathrm{\nu, sca}$ (dashed line) per dust mass for $\lambda=1.14\:\mathrm{mm}$ as the function of maximum grain radius $a_\mathrm{max}$. 
The gray line indicates the radius at which the particle size of the dust becomes comparable to the wavelength, i.e., $2\pi a_{\rm max}=\lambda$, representing the transition point in optical properties.
}
\label{OPtable_1.14mm}
\end{center}
\end{figure}

As an example of our opacity model, Figure\:\ref{OPtable_1.14mm} shows the absorption opacity and effective scattering opacity at $\lambda=1.14\:\mathrm{mm}$, as functions of the maximum grain radius $a_{\rm max}$.
The absorption opacity remains almost constant while the maximum grain radius is below $\sim100\:\mathrm{\mu m}$.
As the grain size exceeds $ a_\mathrm{max} \sim \lambda/(2\pi)$ (the gray line in Figure\:\ref{OPtable_1.14mm}), the absorption opacity rises by approximately five times.
For larger grain sizes, the absorption opacity decreases.

At $0.1\:\mathrm{\mu m}$, the effective scattering opacity is significantly lower than the absorption opacity.
However, the scattering opacity sharply increases as the grain size increases, exceeding the absorption opacity at $a_{\rm max}\gtrsim 100\mathrm{\:\mu m}$.
The scattering opacity peaks at around $300\:\mathrm{\mu m}$ and decreases as the grain size increases further.

Disk temperature calculations require the Rosseland-mean opacity,
which we evaluate from the frequency-dependent opacities obtained from {\tt Optool}:
\begin{equation}\label{eq:kappa}
\cfrac{1}{\kappa_\mathrm{R}(a_\mathrm{max},T)}=
\cfrac{\displaystyle\int^\infty_0 (\kappa_{\nu,\mathrm{abs}} + \kappa^\mathrm{eff}_{\nu,\mathrm{sca}})^{-1} \cfrac{\partial B_\nu(T)}{\partial T} d \nu}
{\displaystyle\int^\infty_0 \cfrac{\partial B_\nu(T)}{\partial T} d \nu}.
\end{equation}
In Equations \eqref{eq:Tam} and \eqref{eq:Taz}, we use the Rosseland-mean opacity $\kappa_\mathrm{R}$ determined by the maximum grain radius $a_\mathrm{max}(r)$ and the disk midplane temperature $T_{\rm mid}(r)$.

\section{Structure and Emission Characteristics of Massive Protostellar Disks with Grain Evolution} \label{sec:Result}
In this section, we first present the physical properties of the constructed disk model\footnote{For a more detailed discussion of the properties of massive protostellar disks with dust evolution, see \cite{Yamamuro2023}.},
followed by the disk emission based on radiative transfer calculations.
The protostellar parameters adopted to the reference model are $M=20\:M_\odot$, $L=4\times10^4\:L_\odot$, $\dot{M}_\mathrm{gas}=1\times10^{-4}\:M_\odot\:\mathrm{yr}^{-1}$, and $r_{\rm disk}=300{\rm\:au}$.
The fragmentation velocity is set to be $v_\mathrm{frag}=1\:\mathrm{m\:s^{-1}}$, which is a commonly adopted value for silicate grains {\citep[e.g.,][]{Blum2000, Guttler10}}.
In the radiative transfer process,
we assume the observing wavelength to be $\lambda=1.14\:\mathrm{mm}$, which value is the same as the previous ALMA Band 6 observations of the GGD27-MM1 disk \citep{Girart2018, Anez+2020}.
For simplicity, we assume a face-on disk orientation ($\mu=1$).

\subsection{Disk Physical Properties}\label{sec:disk_property}

\begin{figure*}
\includegraphics[width=\hsize]{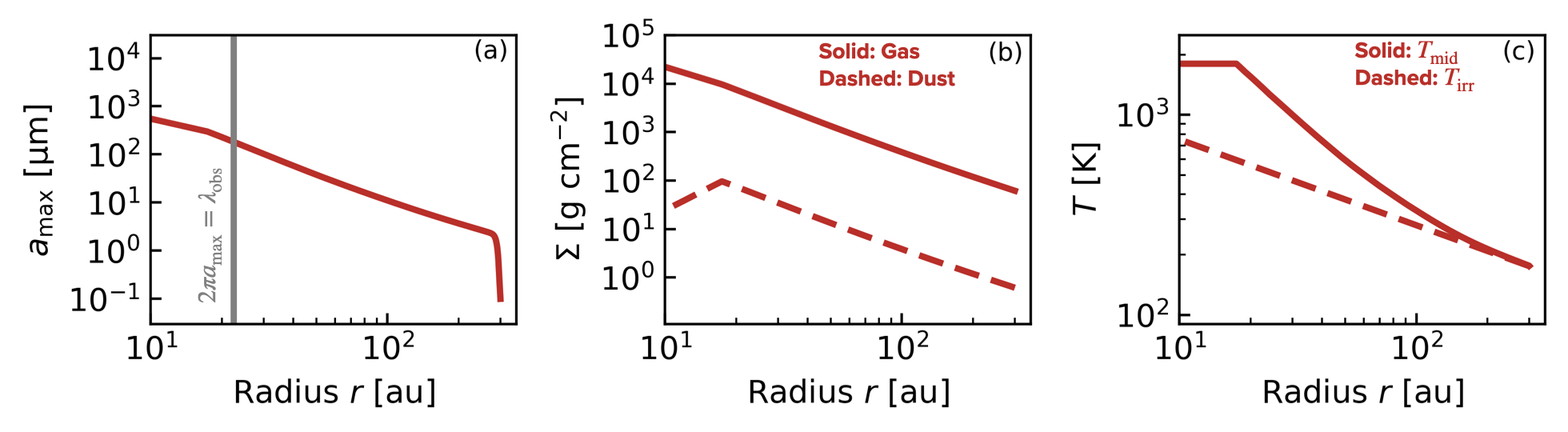}\
\caption{
Radial profiles of physical quantities in the reference model :
(a) maximum {grain radius} $a_{\mathrm {max}}$; 
(b) surface densities of gas $\Sigma_\mathrm{gas}$ (solid) and dust $\Sigma_{\mathrm {dust}}$ (dashed);
(c) midplane temperature $T_\mathrm{mid}$ (solid) and the surface temperature $T_{\rm irr}$ (dashed).
{The gray line in (a) indicates where the maximum grain size reaches the critical condition $2\pi a_\mathrm{max} = \lambda_\mathrm{obs}$, with $\lambda_\mathrm{obs} = 1.14\:\mathrm{mm}$.}
}
\label{fid_disk}
\end{figure*}

Figure \ref{fid_disk} displays the key aspects of the reference model.
The panel (a) shows the maximum grain radius $a_{\mathrm{max}}$, as a function of the radial distance.
At the disk outer edge of $r_{\rm disk}=300\:\mathrm{au}$ where small grains are supplied from the envelope,
the grains undergo local growth and reach of $\gtrsim1{\mathrm{\:\mu m}}$.
This growth is limited by collisional fragmentation, as the collision velocity $\Delta v$ reaches the fragmentation velocity $v_{\rm frag}$.
As the grains accrete inward, they maintain the size determined by the balance between coagulation and fragmentation, i.e., $\Delta v\approx v_{\rm frag}$.
The resulting Stokes number of the grains remains as small as $\sim 10^{-5}$, nearly independent of $r$.

Figure \ref{fid_disk}(b) represents the gas and dust surface densities.
Both the surface densities increase inward, approximately following a single power law of $r^{-1.5}$.
The entire disk is marginally unstable with $Q\sim1$, inducing a strong GI torque characterized by $\alpha\sim0.02$--$0.4$.
Given the small grain size, the GI-induced turbulence is strong ($\alpha \gg \rm St$), resulting in well-coupled gas and dust that accrete toward the central star at the same velocity.
Consequently, the dust-to-gas surface density ratio remains nearly constant across most of the disk, i.e., $\Sigma_{\rm dust }/\Sigma_{\rm gas} \approx f_\mathrm{ dg, ISM}=0.01$.
Additionally, the strong gas-dust coupling prevents dust grains from settling in the vertical direction of the disk \citep{Dubrule1995}.
At the innermost region of $r\lesssim 20\:\mathrm{au}$, the dust surface density decreases due to dust sublimation.

Figure \ref{fid_disk}(c) shows the midplane temperature $T_\mathrm{mid}$ and the irradiation-dominated temperature $T_\mathrm{irr}$, which corresponds to the disk surface temperature.
In the outer region of $r \gtrsim 100{\rm\:au}$,
stellar irradiation dominates the heating, keeping the midplane temperature close to the surface temperature, i.e., $T_\mathrm{mid} \approx T_{\rm irr}$.
In the inner region of $r \lesssim 100{\rm\:au}$,
the midplane temperature becomes significantly higher than the surface temperature,
 due to enhanced accretion heating driven by the deeper gravitational potential and higher optical depth.
At the innermost region around $r\lesssim20~\mathrm{au}$, the midplane temperature reaches the dust sublimation threshold of $T_\mathrm{sub}=1800~\mathrm{K}$, leading to a decline in dust surface density in this region (see panel (b)).
We note that the entire disk maintains temperatures above the ice sublimation threshold of $\sim200{\rm\:K}$,
consistent with the assumption that water ice has sublimated, leaving silicate-dominated dust grains.

\subsection{(Sub-)Millimeter Emission Properties} \label{sec:emission}

\begin{figure*}[t]
\begin{center}
\includegraphics[width=\hsize]{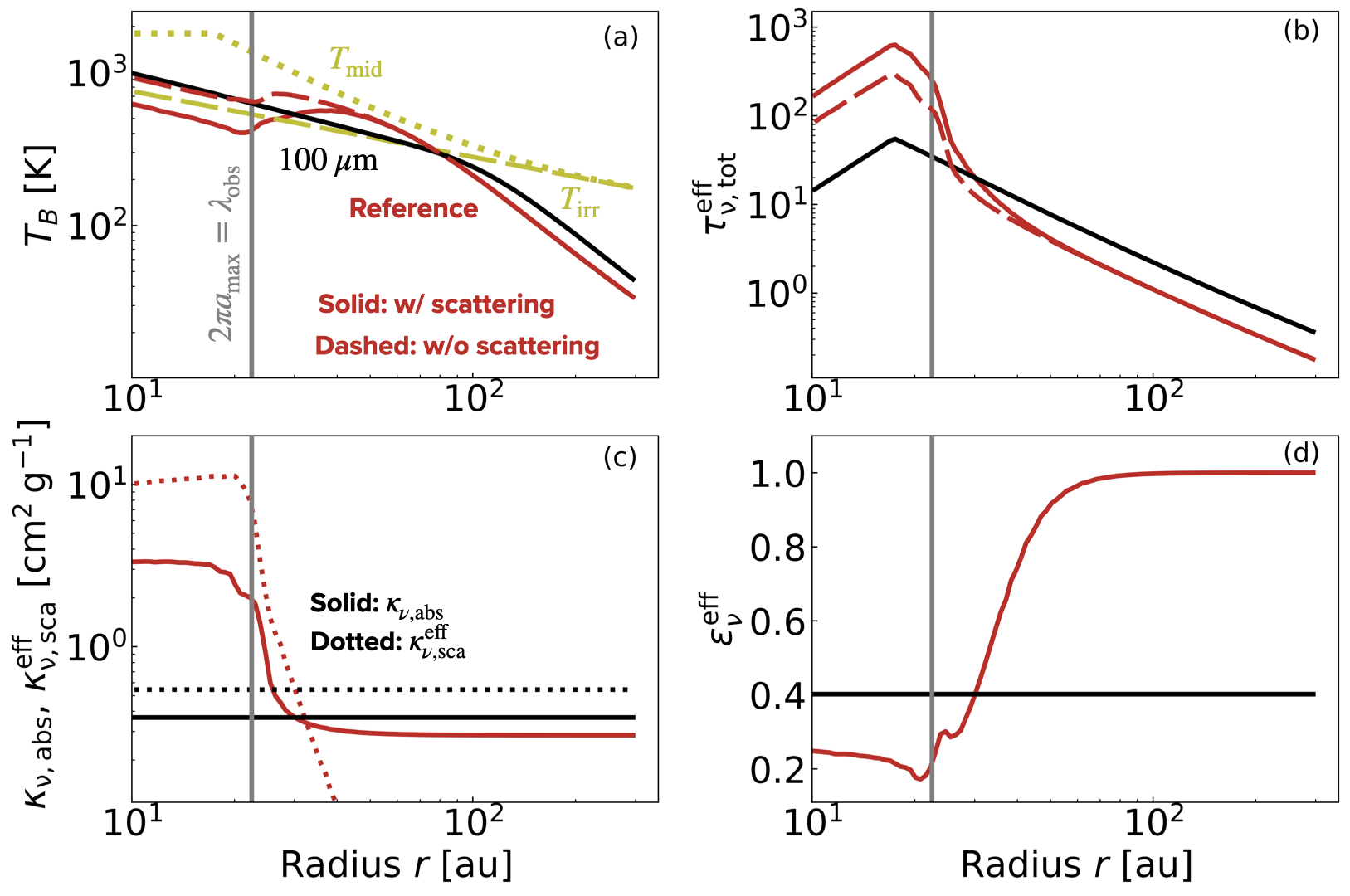}\
\caption{Radial profiles of physical quantities about radiative transfer results of the reference model (solid red lines), the no scattering model (dashed red lines), and {the model with a fixed maximum grain size of $a_\mathrm{max}=100\:\mathrm{\mu m}$ (black lines).}
(a) The brightness temperature $T_\mathrm{B}$;
(b) the effective optical depth $\tau^\mathrm{eff}_\mathrm{\nu,tot}$;
(c) dust absorption (solid) and scattering (dashed) opacity per dust mass $\kappa_\nu$;
(d) absorption efficiency $\epsilon^\mathrm{eff}_\mathrm{\nu}$.
 In panel (a), the yellow dashed lines represent the irradiation temperature $T_\mathrm{irr}$, while the yellow dotted lines represent the midplane temperature $T_\mathrm{mid}$.
 {The gray lines indicate where the maximum grain size reaches the critical condition $2\pi a_\mathrm{max} = \lambda_\mathrm{obs}$, with $\lambda_\mathrm{obs} = 1.14\:\mathrm{mm}$.}}
\label{rad_nog}
\end{center}
\end{figure*}

We now present the dust continuum emission from the reference model at $\lambda=1.14\:\mathrm{mm}$ ($\nu=263{\rm\:GHz}$).
To illustrate the influences of dust optical properties and the grain size growth,
we also compare the reference model with two variations: one without scattering and the other with a constant grain size.
Figure \ref{rad_nog} shows the following properties for these three models:
the brightness temperature $T_\mathrm{B}=I_\nu c^2/(2k_\mathrm{B}\nu^2)$ {in the Rayleigh-Jeans limit},
the effective optical depth $\tau^\mathrm{eff}_\mathrm{\nu,tot}=\sqrt{\kappa_\mathrm{\nu,abs}(\kappa_\mathrm{\nu,abs}+\kappa^\mathrm{eff}_\mathrm{\nu,sca})} \Sigma_\mathrm{dust}$,
the absorption and scattering opacities $\kappa_\mathrm{\nu,abs}$ and $\kappa^\mathrm{eff}_\mathrm{\nu,sca}$,
and the absorption efficiency $\epsilon^\mathrm{eff}_\nu$ as functions of $r$.

In the reference model, the brightness temperature increases inward at $r\gtrsim40{\rm au}$ (Figure \ref{rad_nog}(a)).
Outside of $100\:\mathrm{au}$, where the disk is optically thin with $\tau^{\rm eff}_{\nu, \rm tot}<1$, the brightness temperature follows $T_\mathrm{B} \approx \kappa_\mathrm{\nu, abs} \Sigma_\mathrm{dust} T$, increasing inward due to rising temperature and surface density.
In the intermediate region between $r\approx40$--$80\:\mathrm{au}$, the disk becomes marginally optically thick as $\tau^{\rm eff}_{\nu,\rm tot}\approx1$--$5$.
Here, the brightness temperature surpasses the surface temperature and approaches the midplane temperature, which is elevated by accretion heating.

The inner region of $r\lesssim40{\rm\:au}$ exhibits a distinct emission profile compared to the outer region.
Here, the inner disk is highly optically thick with $\tau^{\rm eff}_{\nu,\rm tot}\gtrsim10$ (panel (b)).
As a result, the effective photosphere at this wavelength (the $\tau_\nu\sim1$ layer from the observer side) shifts from the hotter midplane to the cooler surface.
This causes the brightness temperature to decline at $r\approx40{\rm\:au}$.
Furthermore, for $r\lesssim30{\rm\:au}$, the brightness temperature falls even below the disk surface temperature $T_{\rm irr}$.
The brightness reaches the local minimum at $r\approx20\:\mathrm{au}$ and then increases,
but remains about $20\:\%$ lower compared to $T_\mathrm{irr}$ in the innermost region.
In this region, scattering becomes significant with $\epsilon^\mathrm{eff}_\mathrm{\nu}\approx 0.2$ (Figure \ref{rad_nog}(d)),
with grains growing beyond the critical size of $1.14{\rm\:mm}/(2\pi)\sim 200\:\mathrm{\mu m}$
(see also Figure \ref{OPtable_1.14mm} and Figure \ref{fid_disk}(a)).
Significant scattering is known to make disk emission dimmer than blackbody radiation, even when the disk is optically thick \citep[e.g.,][]{Radiative1986, Sierra2019,Zhu2019,Ueda2020}.

To better illustrate the scattering-induced dimming, we also perform the disk emission calculation without scattering, i.e., with the absorption efficiency fixed at $\epsilon^{\rm eff}_\nu=1$ (see the dashed red line of Figure \ref{rad_nog}(a)).
In the outer region of $r\gtrsim40\:\mathrm{au}$,
the brightness profile of the no-scattering model follows a similar trend to that in the reference model.
However, within this radius, where scattering significantly affects the emission in the reference model, the no-scattering model appears brighter.
Specifically, $T_{\rm B}$ of the no-scattering model remains above the disk surface temperature even at the innermost region,
whereas it falls below $T_{\rm irr}$ in the reference model.

In the reference model, the grain size increases toward the central star.
To examine the effect of this grain evolution on disk emission, we perform an additional transfer calculation where the $1.14{\rm\:mm}$ opacities are fixed to those for $a_\mathrm{max}=100\: \mathrm{\mu m}$ across the entire disk (see the black line in Figure \ref{rad_nog} (a)).
The thermal and density profiles are maintained from the reference model.
In the optically-thin outer region beyond $r\gtrsim100\:\mathrm{au}$, where the grain size in the reference model is smaller than $100{\rm\:\mu m}$,
the emission is more pronounced in this no-coagulation model compared to the reference case (Figure \ref{rad_nog}(a)).
This occurs because the opacities are higher in this region than in the reference model (Figure \ref{rad_nog}(c)).
The absorption efficiency is $\epsilon^\mathrm{eff}_\nu\approx0.4$ in the no-coagulation model (Figure \ref{rad_nog}(d)),
indicating that the scattering effect is moderate but not particularly strong throughout the disk.
In the intermediate region around $r\approx30$--$80\:\mathrm{au}$, this moderate scattering dimming, combined with high optical depth, results in a lower $T_{\rm B}$ for the no-coagulation model compared to the reference model.
However, in the innermost region around $r\approx20$--$30\:\mathrm{au}$,
where the reference model dims below $T_{\rm irr}$ due to significant scattering ($\epsilon^\mathrm{eff}_\nu\approx0.2$),
the no-coagulation model does not exhibit such an emission drop, maintaining brightness slightly above the disk surface temperature.
These results highlight that the emission drop in the inner disk is a distinctive feature of grain size evolution, particularly as grains grow beyond the critical size of $\lambda/(2\pi)$ along the radial direction.

\section{Modeling the GGD27-MM1 Disk}\label{sec:bestfit}
Using our disk model and radiative transfer calculations, we aim to reproduce the observed radial profile of the $1.14$-$\mathrm{mm}$ continuum emission from the massive protostellar disk around GGD27-MM1, as reported by \cite{Anez+2020}.
Specifically, we seek for models that satisfy the constraints from polarization observations \citep{Girart2018} on grain size at $r=100\:\mathrm{au}$.

We note that \cite{Anez+2020}, who reported the $1.14$-$\mathrm{mm}$ emission profile, also performed model fitting to reproduce the observed features of the GGD27-MM1 disk.
Key differences between our model and that of \cite{Anez+2020} lie in the treatment of grain behavior, specifically in size growth along the radial direction and vertical settling.
\cite{Anez+2020} parameterized the maximum grain size uniformly across the disk and identified a best-fit model with a maximum size of $500\:\mathrm{\mu m}$.
In contrast, our model self-consistently calculates the maximum grain radius by solving for grain growth and fragmentation, resulting in a radially varying grain size distribution.
Additionally, \cite{Anez+2020} assumed that larger grains settle to 0.1 times the gas scale height.
In our disk model, grain settling can occur when the grain size becomes sufficiently large \citep[see Equation 27 of][]{Yamamuro2023}.
However, due to the high gas surface density and strong turbulence in the disk, settling is effectively suppressed (i.e., the Stokes number is much smaller than the $\alpha$ value).
While vertical settling is significant in protoplanetary disks with low gas surface density and weak turbulence, it is unlikely to occur in massive protostellar disks like GGD27-MM1.
{The near-far side asymmetry in polarization observed in the GGD27-MM1 disk \citep{Girart2018} is consistent with the interpretation that the disk is geometrically thick and remains largely unsettled \citep{Yang2017}.}
These differences in the treatment of grain behavior affect the (sub-)mm continuum emission from disks.
We will further discuss the differences in the resulting continuum properties in Section \ref{sec:Dis}.

\subsection{Methodology for Parameter Search}\label{fitting_pro}
Our goal is to identify a model that reproduces the radial profile of the $1.14\:\mathrm{mm}$ emission of the GGD27-MM1 disk \citep[Figure 7 of][]{Anez+2020}.
Our model involves three key parameters: fragmentation velocity $v_{\rm frag}$, accretion rate $\dot{M}_{\rm acc}$, and disk radius $r_{\rm disk}$.
We identify candidate models by minimizing the normalized mean squared error (NMSE) between the observed and modeled intensity profiles.
The stellar mass is fixed at $M=20\:M_\odot$, based on a dynamical analysis of \ce{SO2} line emissions \cite{Anez+2020}.
The stellar luminosity is set at $L=1.4\times10^4\:L_\odot$, estimated from near, mid and far -infrared observations \citep{Yamashta1987}.

As we will show in Section~\ref{sec:Bestfit}, this fiducial approach struggles to reproduce the observed brightness distribution in the inner region, where significant scattering reduces the modeled brightness below observed levels.
To address this discrepancy, we extend the parameter exploration with two alternative scenarios in Section \ref{43}.
In Section \ref{sec_limit}, we introduce a size limit on the maximum grain size, $a_\mathrm{lim}$.
This size limitation could occur if colliding grains neither coagulate nor fragment but instead bounce off each other and prevent further growth,
a phenomenon known as the bouncing barrier in planetesimal formation studies \citep[e.g.,][]{Guttler10,Zsom2010, Oshiro2025}.
In Section \ref{Lumino_fit}, we explore the impact of uncertainties in the protostellar luminosity by treating the stellar luminosity as a free parameter.
All explored parameters are summarized in Table \ref{c2t_table_all}.

In the calculations of the disk structure, we set the disk radius to $r_\mathrm{disk} = 300{\rm\:au}$ for all models.
This simplification is based on the fact that silicate grains near the outer edge locally grow to the fragmentation-limited size, and thus the physical properties of the disk structures remain largely unaffected by the choice of $r_\mathrm{disk}$ (see Figure \ref{fid_disk}(a)).
In contrast, during the post-processing radiative transfer calculations, $r_\mathrm{disk}$ is treated as a free parameter to explore its impact on the observed intensity profiles, with $I_\nu = 0$ imposed for $r > r_\mathrm{disk}$.

To compare the model with the observations of the GGD27-MM1 disk,
we project the model profile onto the plane of the sky, incorporating the disk inclination \citep[$\mu = \cos 49^\circ$;][]{Anez+2020}.
The two-dimensional intensity profile is expressed as $I_\nu(x,y)=I_\nu(r=\sqrt{x^2+y^2/\mu^2})$,
where the $x$-$y$ coordinate system is set in the sky plane, with $x$ aligned to the major axis of the projected disk.
To account for the finite spatial resolution of the observations, we apply Gaussian smoothing:
\begin{equation}
\begin{split}
    &{\overline{I_\nu}(x,y)}= \frac{1}{{{2\pi} \sigma_{\rm sm}^2}}\\
    &\times \int\!\!\!\int I_\nu(x', y') \,
        \exp\left[
            -\frac{(x-x')^2 + (y-y')^2}{2 \sigma_{\rm sm}^2}
        \right] dx' dy',
\end{split}
\end{equation}
where ${\overline{I_\nu}(x,y)}$ represents the smoothed intensity, and $\sigma_{\rm sm}$ is the smoothing standard deviation.
For simplicity, our Gaussian smoothing uses a radially symmetric beam,
ignoring the elliptical shape and position angle of the actual observation beam.
The source distance is $d = 1.4{\rm\:kpc}$,
and the observation beam has a full width at half maximum (FWHM) of $\sqrt{45.0{\rm\:mas} \times 38.3{\rm\:mas}}$ \citep{Anez+2020}.
Thus, the smoothing standard deviation is set to $\sigma_{\rm sm}=d\:{\rm FWHM}/(2\sqrt{2\log2})=58{\rm\:au}$.
The intensity for radii $r < 10{\rm\:au}$, which lies inside the calculation domain, is approximated by the value at $r = 10{\rm\:au}$.

We utilize observational data from \cite{Anez+2020}, specifically Figure 7,
which presents the radial profile of dust continuum intensity observed at $1.14{\rm\:mm}$.
This profile includes 16 data points spanning a radial distance of $r = 0$--$220{\rm\:au}$, denoted by $I_{\nu,\mathrm{obs}}(r_k)$, where $k = 1, 2, \dots, 15$.
To assess how well each model matches the observations,
we use the Normalized Mean Squared Error (NMSE) metric, which provides a quantitative measure of the agreement between the modeled and observed intensity profiles:
\begin{equation} \label{eq:in}
    \mathrm{NMSE} = \frac{\displaystyle\sum_{k=1}^{15} \left( \overline{I_{\nu}}(r_k) - I_{\nu,\mathrm{obs}}(r_k) \right)^2}{\displaystyle\sum_{k=1}^{15} \left( I_{\nu,\mathrm{obs}}(r_k) \right)^2}.
\end{equation}
We analyze the model profile only along the disk major axis,
i.e., $\overline{I_{\nu}}(r_k)=\overline{I_{\nu}}(x=r_k,y=0)$.
The observed point at $r=r_0$ is excluded from the NMSE calculation because our disk model does not cover this innermost region.
We perform a parameter survey to identify models that minimize the NMSE,
thereby achieving the best match with the observed radial profile of the GGD27-MM1 disk.

\begin{figure}[htbp]
\begin{center}
\includegraphics[width=\hsize]{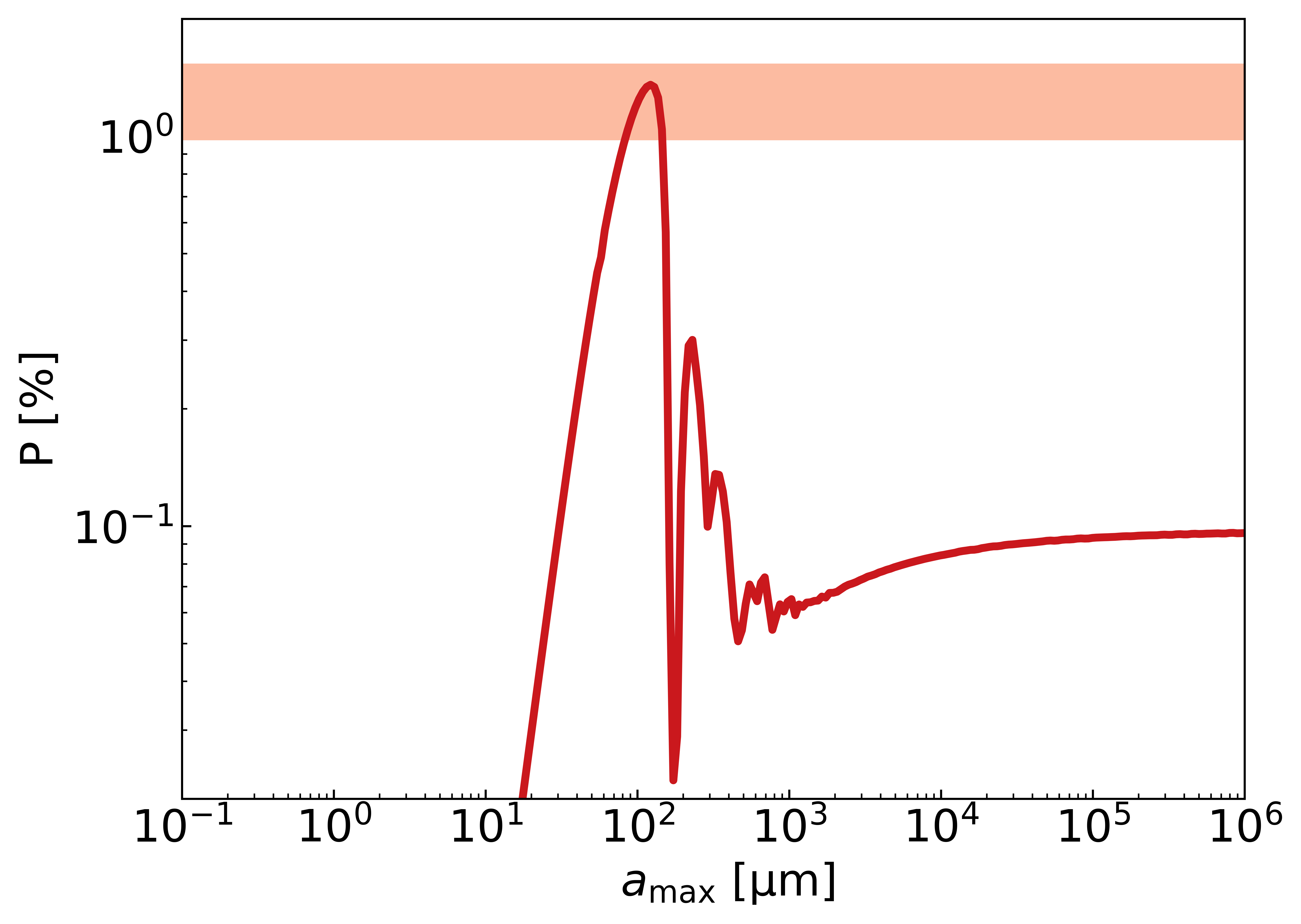}\
\caption{The expected polarization degree at the wavelength of $1.14\:\mathrm{mm}$ as a function of the maximum grain radius, $a_\mathrm{max}$.
The shaded region indicates the observed polarization range of $1$--$1.5\:\%$.
}
\label{fig:Polpl}
\end{center}
\end{figure}

In addition to the overall continuum profile, we aim to constrain the grain size that reproduces the observed polarization features of the GGD27-MM1 disk \citep[see also Section 5.2 of][]{Yamamuro2023}.
In the {southwestern} region of the disk at $r \approx 100{\rm\:au}$,
the polarization vectors align with the disk minor axis with polarization degrees of  $P=1$--$1.5\:\%$, likely caused by self-scattering of grown grains \citep{Girart2018}.
Using the method of \cite{Kataoka2015}, we estimate the grain size that matches these polarization features.
The polarization degree is evaluated as:
\begin{equation}
    P=- C \frac{Z_\mathrm{12}}{Z_\mathrm{11}} \frac{\kappa^\mathrm{eff}_\mathrm{\nu,sca}}{\kappa_\mathrm{\nu, abs} + \kappa^\mathrm{eff}_\mathrm{\nu,sca}},
\end{equation}
where  $C$ is a constant factor related to scattering efficiency, and $Z_\mathrm{11}$ and $Z_\mathrm{12}$ are elements of the scattering matrix {evaluated at a scattering angle of $90^\circ$}.
Following \cite{Kataoka2015}, we adopt $C = 2\:\%$.
We use {\tt Optool} to calculate the scattering matrix elements for astronomical silicate.
Figure \ref{fig:Polpl} shows the calculated polarization degree as a function of the maximum grain radius.
We find that grains with sizes in the range of $a_{\rm max}=100$--$150{\rm\:\mu m}$ reproduce the observed value of $P=1$--$1.5{\:\%}$ at $r \approx 100~\rm au$.
Therefore, we focus on models with the smallest NMSE among those that satisfy this grain size constraint.
{Here, we defined the grain size constraint such that grain with $a_\mathrm{max}=100$--$150\:\mathrm{\mu m}$ exists within the radial range $r=90$--$110\:\mathrm{au}$, corresponding to $\pm10\:\%$ around $100\:\mathrm{au}$. }

\subsection{Fiducial Case}
\label{sec:Bestfit}
To identify candidate models capable of reproducing the $1.14\:\mathrm{mm}$ observations of the GGD27-MM1 disk,
we first systematically explored three key parameters: $\dot{M}_\mathrm{gas}$, $r_\mathrm{disk}$, and $v_\mathrm{frag}$ (Table \ref{c2t_table_all}).
Here, we present the model that achieves the lowest NMSE for the intensity profile while also satisfying the grain size constraint.
The parameters of this model are $\dot{M}_\mathrm{gas}=1\times10^{-3}\:M_\odot \mathrm{yr}^{-1}$, $r_\mathrm{disk}=150\:\mathrm{au}$, and $v_\mathrm{frag}=14\:\mathrm{m\:s^{-1}}$ (Table \ref{table_fit}).

\begin{figure}[htbp]
\begin{center}
\includegraphics[width=\hsize]{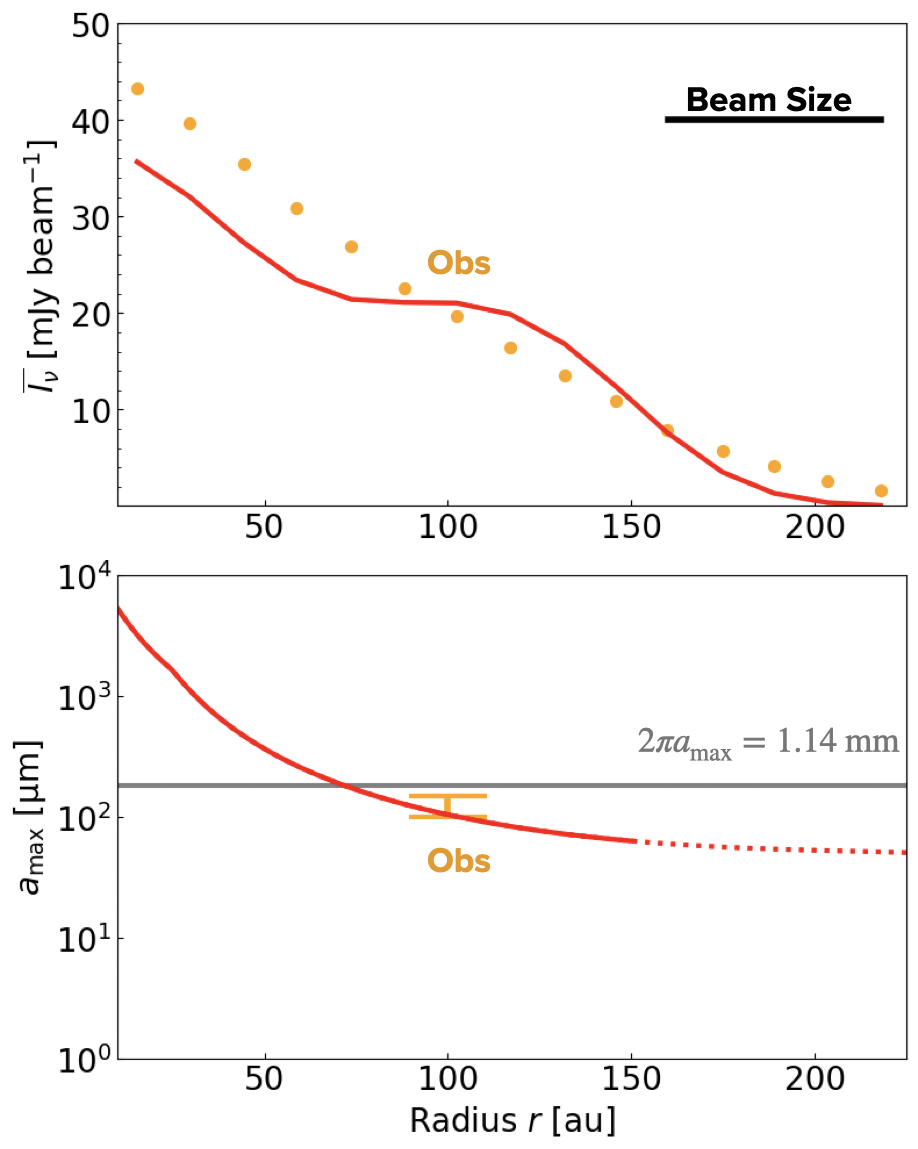}\
\caption{The fitted model. Upper panel: The $1.14\:\mathrm{mm}$ continuum emission profiles of the fiducial fitted model (red line) are compared with the observed data (orange points; \citet{Anez+2020}).
The black line in the upper right part of the panel represents the beam size.
Lower panel: the maximum grain radius in the fiducial fitted model.
The vertical bar denotes the inferred range for the maximum grain radius of $a_\mathrm{max}=100$--$150\:\mathrm{\mu m}$ at $100\:\mathrm{au}$, based on the assumption that polarized emission from this region is attributable to dust self-scattering.
{The dotted line indicates the maximum grain radius in the region beyond $r_\mathrm{disk}$, which was excluded from the post-processing radiative transfer calculations.}
}
\label{Best_fit}
\end{center}
\end{figure}

The top panel of Figure \ref{Best_fit} compares the intensity profiles of the model (red line) and the observations of GGD27-MM1 \citep[orange points, adapted from][]{Anez+2020},
while the bottom panel shows the {maximum grain size}.
The model satisfies the grain size constraint of $a_{\rm\:max}=100$--$150{\rm\:\mu m}$ at $100{\rm\:au}$, with grains continuing to grow inward.
However, unlike the observed intensity profile, the model exhibits the distinct feature:
the intensity does not increase smoothly toward the center but instead flattens or slightly declines in the inner region.
As discussed in Section \ref{sec:emission}, this behavior arises from significant grain growth,
with sizes exceeding the critical threshold of $\lambda / (2\pi) \approx 200\:\mathrm{\mu m}$.
The large grains cause the inner region to become significantly optically thick at $\lambda \approx 1~\rm mm$, preventing the deeper, hotter interior to contribute to the emergent intensity.
Furthermore, significant scattering dims the emission, leading to a further reduction in the observed brightness temperature.
Consequently, the model intensity is approximately $20$--$30\:\%$ lower than the observed values in the inner region of $r \lesssim 50\:\mathrm{au}$, despite achieving the smallest $\mathrm{NMSE}$ among the wide range of surveyed parameters.
This trend appears in all models from the fiducial three-parameter survey,
where the inner region consistently exhibits intensities that are more than $20\:\%$ lower than the observed values (Appendix \ref{app:fid}).
This discrepancy is attributed to the grain radius exceeding the critical size of $\approx200\:\mathrm{\mu m}$ in the inner disk,
leading to increased opacity and insufficient reproduction of the GGD27-MM1 observations.

\begin{deluxetable*}{lccccc}
\tablecaption{Explored Parameters for Modeling the GGD27-MM1 Disk}
\label{c2t_table_all}
\tablehead{
\colhead{}&\colhead{$L\:(L_\odot)$}&\colhead{$\dot{M}_{\rm gas}$\:($M_\odot\:\mathrm{yr}^{-1}$)} &\colhead{$r_{\rm disk}$\:(au)} & \colhead{$v_{\rm frag}$\:($\rm m\:s^{-1}$)}&\colhead{$a_\mathrm{lim}$\:($\mathrm{\mu m}$)}
}
\startdata
Fiducial Case  & $1.4\times10^4$ & $10^{-5}$--$10^{-3}$ & $150$--$200$ & $0.1$--$30$ & - \\
Grain Size Limit & $1.4\times10^4$ & $10^{-5}$--$10^{-3}$ & $150$--$200$ & $0.1$--$30$ & $100$--$200$ \\
High Luminosity  & $1$--$7\times10^4$ & $10^{-5}$--$10^{-3}$ & $100$--$200$ & $0.1$--$30$ & - \\
~~~~~~~~~~Step & $\Delta L = 10^4 $ & $\Delta \mathrm{log} \dot{M}_\mathrm{gas}=\mathrm{log}(10^{-3}/10^{-5})^{1/20}$ & $\Delta r_\mathrm{disk}=10$ & $\Delta \mathrm{log}\:v_\mathrm{frag} = \mathrm{log}(30/0.1)^{1/30}$ \\
\enddata
\tablecomments{We fix the values of the stellar mass at $M=20\:M_\odot$, grain density at $\rho_{\rm int}=3.3{\rm\:g\:cm^{-3}}$, and the interstellar dust-to-gas mass ratio at $f_{\rm dg, ISM}=0.01$ for all models.}
\end{deluxetable*}

\begin{deluxetable*}{lcccccc}
\tablecaption{Optimal Parameter Sets for Reproducing the GGD27-MM1 Disk}
\label{table_fit}
\tablehead{
\colhead{}&\colhead{$L\:(L_\odot)$}&\colhead{$\dot{M}_{\rm gas}$\:($M_\odot\:\mathrm{yr}^{-1}$)} &\colhead{$r_{\rm disk}$\:(au)} & \colhead{$v_{\rm frag}$\:($\rm m\:s^{-1}$)}&\colhead{$a_\mathrm{lim}$\:($\mathrm{\mu m}$)} & \colhead{$\mathrm{NMSE}$}}
\startdata
Fiducial Case & $1.4\times10^4$ & $1.0\times 10^{-3}$ & $150$ & $14$ & - & {$2.4\times10^{-3}$}\\
Grain Size Limit & $1.4\times10^4$ & $5.0\times10^{-4}$ & $160$ & $12$ & $160 $  &{$5.5\times10^{-4}$}\\
High Luminosity & $7.0\times10^4$ & $7.9\times 10^{-4}$ & $150$ & $17$ & -  &{$4.6\times10^{-4}$}\\
\enddata
\tablecomments{These values are presented with two significant figures.}
\end{deluxetable*}

\subsection{Exploring Alternatives to Explain the Disk Brightness}\label{43}
Here, we introduce modifications to the model to better reproduce the intensity profile in the inner region of the disk.

\subsubsection{Grain Size Limit} \label{sec_limit}
In the models of the fiducial parameter survey,
the maximum grain size continues to increase inward across the entire disk (Figure \ref{Best_fit}).
This behavior is a general consequence of the fragmentation-limited growth with constant sticking threshold velocity ($v_{\rm frag}$ in this case) \citep[e.g.,][]{Birnstill2023}.
However, the maximum grain size would be more insensitive to $r$ if the sticking threshold velocity decreases with increasing grain size. 
Such a sticking threshold is often observed when grain growth is limited by bouncing collisions \citep{Guttler10, Arakawa2023, Oshiro2025}\footnote{We note, however, that whether bouncing indeed yields a radially nearly constant maximum grain size depends on the details of the turbulence-induced collision velocity \citep{Dominik2024}.}.
Here, we explore the possibility that bouncing or any other unknown growth barrier inhibits grain growth beyond a size limit $a_{\rm lim}$.
By treating $a_{\rm lim}$ as an additional model parameter alongside the three fiducial ones (Table \ref{c2t_table_all}),
we re-explore the parameter space to identify models that minimize the NMSE for the observed brightness profile of GGD27-MM1.

\begin{figure}[htbp]
\begin{center}
\includegraphics[width=\hsize]{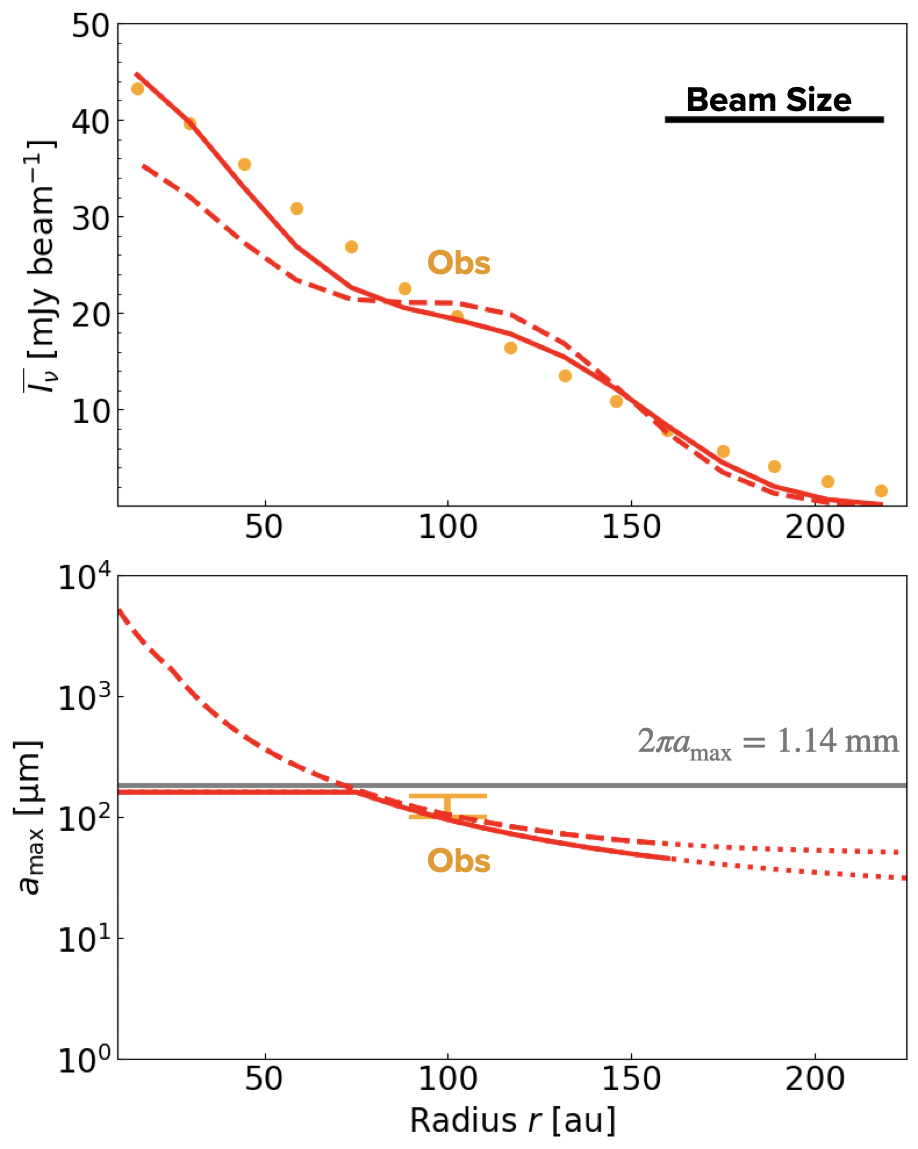}\
\caption{ The same as Figure \ref{Best_fit}, but for the model with a grain size limit at $a_{\rm lim} = 160~\rm \mu m$ (see Section \ref{sec_limit}).
{The dashed lines represent the results shown in Figure \ref{Best_fit}, included for comparison.}
}
\label{GGD_stop}
\end{center}
\end{figure}

We find that the parameter set for the revised model with the bouncing barrier is
$\dot{M}_\mathrm{gas}=5.0\times10^{-4}\:M_\odot \:\mathrm{yr}^{-1}$, $r_\mathrm{disk}=160\:\mathrm{au}$, $v_\mathrm{frag}=12\:\mathrm{m\:s^{-1}}$,
and $a_\mathrm{lim}=160\:\mathrm{\mu m}$.
Figure \ref{GGD_stop} shows the radial profiles of the intensity and maximum grain size for the revised model.
The intensity profile shows a significant improvement in the inner disk ($r< 100\:\mathrm{au}$),
where the discrepancy with observations is reduced to an average of $8\:\%$, compared to the $20$–$30\:\%$ mismatch in the fiducial case.
The improved agreement is attributed to the grain size limit of $a_{\rm lim}=160{\rm\:\mu m}$, which remains below the critical threshold of $\lambda/(2\pi)$.
This leads to two key effects:
(1) lower total opacity ($\kappa_{\nu,\mathrm{abs}} + \kappa_{\nu,\mathrm{sca}}^{\rm eff}$), allowing radiation from the hotter midplane to escape more effectively,
and (2) higher absorption efficiency ($\epsilon_{\nu}^{\rm eff}$), reducing the impact the scattering-induced dimming.
This model thus offers one possible solution to the excessive dimming issue identified in the fiducial case.

\subsubsection{Higher Luminosity} \label{Lumino_fit}
Next, we explore another alternative scenario invoking a higher stellar luminosity.
If the stellar luminosity is higher, the disk surface temperature increases,
potentially alleviating the dimming issue of the inner-disk brightness seen in the fiducial case.

So far, we have assumed a stellar luminosity of $1.4 \times 10^4\: L_\odot$, based on infrared spectral energy distribution (SED) \citep{Yamashta1987}.
However, the observed SED shifts toward longer wavelengths compared to direct stellar radiation because it is primarily re-emitted from the surrounding dusty envelope.
Additionally, the GGD27-MM1 disk is also associated with jets HH80--81, creating outflow cavities \citep[cf.][]{Bally2023}.
These non-spherical structures around protostars lead to the ``flashlight effect'',
where radiation escapes preferentially along the cavity \citep[e.g.,][]{Nakano1995, Yorke1999}.
Radiative transfer modeling by \cite{Zhang2014} demonstrates that
this effect can lead to a stellar luminosity underestimated by a factor of $2$--$5$ when the luminosity estimate relies solely on integrated infrared SEDs.
Moreover, given the stellar mass of $20\:M_\odot$ and the accretion rate of $\gtrsim10^{-4}\:M_\odot{\rm\:yr^{-1}}$,
protostellar models suggest that the stellar luminosity could plausibly reach nearly $10^5\:L_\odot$ \citep[e.g.,][]{Hosokawa2009}.
To examine the high-luminosity scenario, we now treat the stellar luminosity as an additional free parameter, exploring the range of $1$--$7 \times 10^4\:L_\odot$ to identify the model with the lowest NMSE (Table \ref{c2t_table_all}).
In this exploration, we do not impose the $a_{\rm lim}$ limit on the maximum grain size.

\begin{figure}[htbp]
\begin{center}
\includegraphics[width=\hsize]{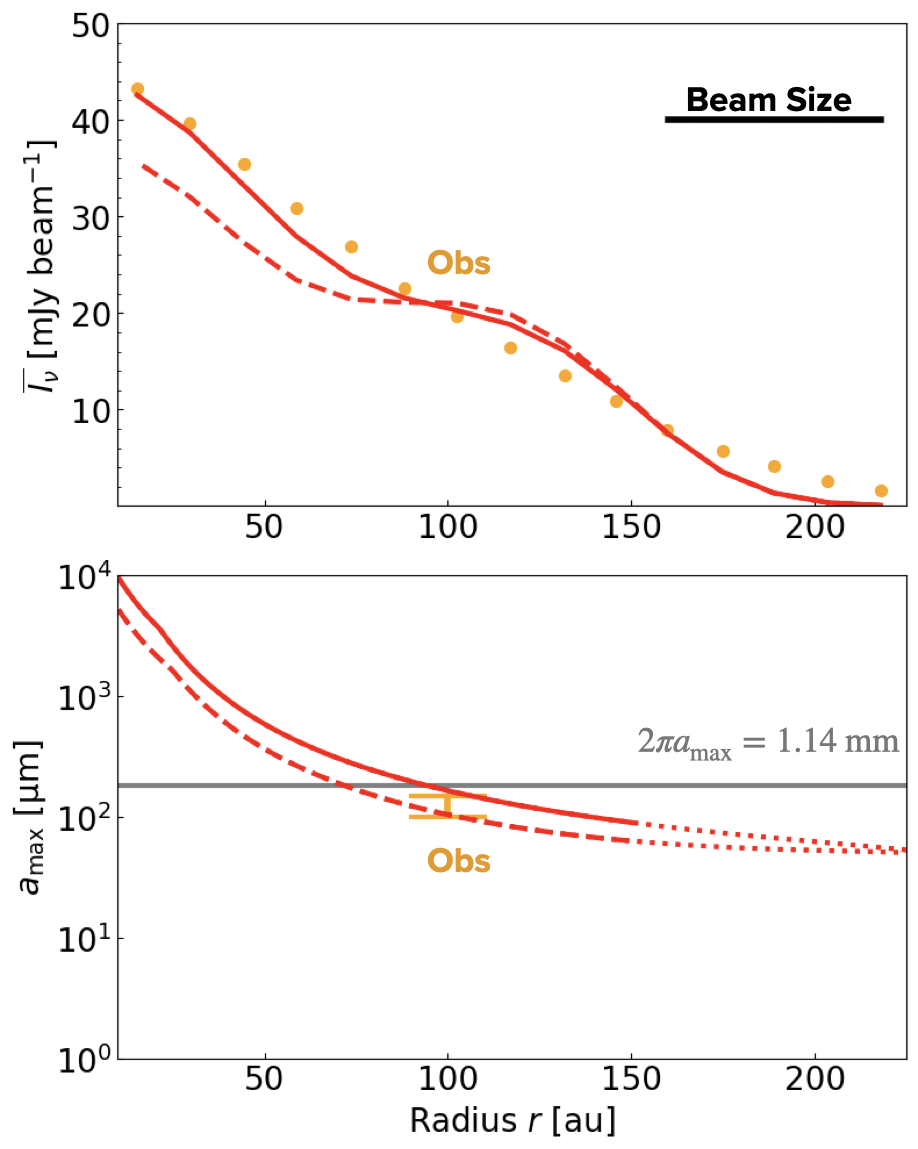}\
\caption{The same as Figure \ref{Best_fit}, but for the model with a higher stellar luminosity of $L=7\times 10^4 L_\sun$ (see Section \ref{Lumino_fit}).
{The dashed lines represent the results shown in Figure \ref{Best_fit}, included for comparison.}
}
\label{Lumino_fig}
\end{center}
\end{figure}

We find that the revised model best matches the observation when
$v_\mathrm{frag}=17\:\mathrm{m\:s^{-1}}$,
$\dot{M}_\mathrm{gas}=7.9\times10^{-4}\:M_\odot \mathrm{yr}^{-1}$,
$r_\mathrm{disk}=150\:\mathrm{au}$, and $L=7\times10^4\:L_\odot$.
The stellar luminosity in this model is five times higher than that in the fiducial case.
Figure \ref{Lumino_fig} shows the radial profiles of the intensity and maximum grain size for this higher-luminosity model.
The discrepancy with the observation in the inner disk ($r< 100\:\mathrm{au}$) is reduced to an average of $6\:\%$ in this model.
In this scenario, grains grow beyond the critical size of $\lambda/(2\pi)$ in the inner region, leading to a scattering-induced dimming effect.
Nevertheless, the higher stellar luminosity prevents a significant drop in the inner-disk brightness, maintaining a closer match to the observed intensity profile.
{In the outer disk, despite the higher stellar luminosity, the flux per beam remains comparable to that of the grain-size-limited model.
This is because the high-luminosity model has a smaller disk radius than the grain-size-limited model (see Table \ref{table_fit}), which results in similar thermal emission from the outer regions.}

\section{Discussion} \label{sec:Dis}
Here, we compare the models derived in the previous section that can reproduce the observations of the GGD27-MM1 disk and explore how hypothetical follow-up observations could help distinguish between these scenarios.
Then, we discuss the broader implications of our findings for rocky planetesimal formation and massive star formation.

\subsection{Comparison of the Plausible Models for the GGD27-MM1 Disk}
\label{sec:multi}
We found two plausible scenarios that reproduce both the $1.14{\rm\:mm}$ emission profile and the grain size constraint of GGD27-MM1: the grain size limit scenario (Section \ref{sec_limit}) and the high-luminosity scenario (Section \ref{Lumino_fit}).
Both scenarios also share similar optimal parameters, except for the grain size limit and stellar luminosity (Table \ref{table_fit}).

A key result across all cases, including the fiducial case, is that the fragmentation velocity consistently falls within the range of $v_\mathrm{frag}=12$--$17{\rm\:m\:s^{-1}}$.
{This result is somewhat larger than the simplified estimate of $\approx10\:\mathrm{m\:s^{-1}}$ presented in \cite{Yamamuro2023}, which was derived based solely on grain size constraints without considering the brightness profile.
In any case, these results provide important insights into the sticking properties of silicate grains, and consequently, the formation of rocky planetesimals (Section \ref{Planet}).}

The gas accretion rates derived from the two scenarios are also nearly identical, at $5$–$7.9\times10^{-4}{\:M_\odot\rm\:yr^{-1}}$.
However, this value is about $7$--$10$ times higher than the best-fit model reported in the previous study \cite{Anez+2020}.
The primary reason for this discrepancy likely lies in
their assumption that larger grains settle to 0.1 times the gas scale height.
Under this settling assumption, large grains with $a_{\rm max} > \lambda/(2\pi)$ are effectively obscured by smaller grains in the upper layers of the disk.
This reduces the impact of scattering-induced dimming and allows the inner-disk brightness to remain high, even with a lower accretion rate.
However, their best-fit model estimates a surface density of $400$--$1400{\rm\:g\:cm^{-2}}$ and a turbulence parameter of $\alpha=0.1$, indicating strong turbulence and a high surface density.
Even for grains in the size range of $100$--$1000{\rm\:\mu m}$,
significant vertical settling is unlikely because the condition ${\rm St}/\alpha \ll 1$ holds in this regime.
Thus, we consider the accretion rate of $\approx7\times10^{-4}{\:M_\odot\rm\:yr^{-1}}$ derived from our models to better represent the physical conditions of the GGD27-MM1 disk.

While the two scenarios share similarities in their $1.14{\rm\:mm}$ emission profiles and some parameters, they also exhibit significant differences.
In the high-luminosity case, the luminosity is five times higher than the grain size limit case, and the maximum grain size is more than an order of magnitude larger in the innermost region ($r < 30{\rm\:au}$).
Determining which scenario more accurately reflects the conditions of the GGD27-MM1 disk is crucial for uncovering its underlying physical properties.
{The polarization map shows a fraction of about $1–1.5\: \%$ that appears to extend slightly inward to within $100\:\mathrm{au}$ \citep[see Figure 3 of][]{Girart2018}. 
However, given the angular resolution of $56\:\mathrm{au}$, it is difficult to determine whether this polarization signal truly originates from within $100\:\mathrm{au}$. For this reason, we consider it premature to judge which scenario is more plausible based solely on the polarization detected in the inner disk.}

\begin{figure*}[ht]
\begin{center}
\includegraphics[width=\hsize]{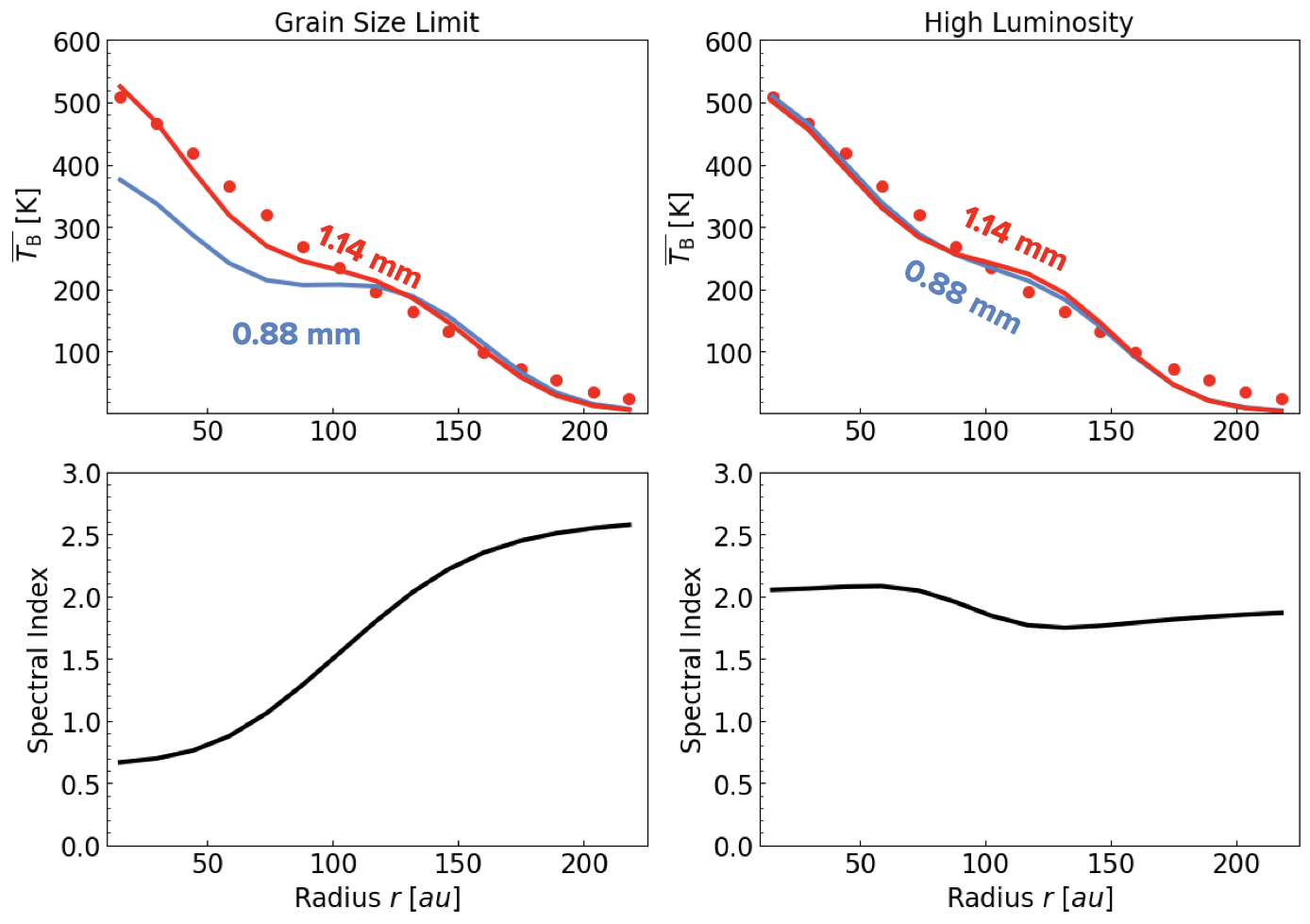}\
\caption{Upper panel: Two-wavelength predictions for two models. Left: brightness temperature profiles at $\lambda=1.14\:\mathrm{mm}$ (red) and $0.88\:\mathrm{mm}$ (blue) for the grain size limit model (Section \ref{sec_limit}). Right: similar to the left panel, but depicting the high-luminosity model (Section \ref{Lumino_fit}). These model projections suggest that future $0.88\:$mm observations could effectively distinguish between these two scenarios. Lower: Spectral index for two models. Radial profiles at $\lambda=0.88-1.14\:\mathrm{mm}$ for the grain size limit scenario (left) and the high-luminosity scenario (right).
}
\label{Multiwave}
\end{center}
\end{figure*}

We propose that follow-up observations at shorter wavelengths will be key to making this distinction.
To illustrate this, we conducted radiative transfer calculations at $\lambda=0.88{\rm\:mm}$ (ALMA Band 7) for the optimal models of each scenario (the same beam size is assumed).
Figure \ref{Multiwave} shows the brightness temperature profiles and spectral indices at $\lambda = 0.88$ and $1.14{\rm\:mm}$ for both scenarios.
The grain size limit model and the high-luminosity model are compared to highlight their distinct emission characteristics.

In the grain size limit case (left panels),
the $0.88{\rm\:mm}$ brightness temperature is approximately $30\:\%$ lower than that at $1.14{\rm\:mm}$ within $100\:\mathrm{au}$.
This difference arises because the shorter wavelength corresponds to higher opacity, causing the observed radiation to originate from cooler upper layers of the disk rather than the accretion-heated midplane.
Additionally, scattering dimming becomes effective at $0.88{\rm\:mm}$
because the grain size limit of $a_{\rm lim} = 160{\rm\:\mu m}$ slightly exceeds the critical size $\lambda/(2\pi)$ for this wavelength.
These combined effects lead to dimmer emission at $0.88{\rm\:mm}$ compared to $1.14{\rm\:mm}$.
The resulting spectral index within $100{\rm\:au}$ is as low as $0.5$--$1.5$,
significantly below the blackbody value of 2.
This behavior reflects the influence of scattering dimming and the disk vertical temperature gradient,
as highlighted in previous studies \citep{Liu2019, Sierra2020}.

On the other hand, in the high-luminosity case (right panels),
the brightness temperatures at $0.88{\rm\:mm}$ and $1.14{\rm\:mm}$ show almost identical profiles.
Without the grain size limit,
the maximum grain sizes within $r \lesssim 100{\rm\:au}$ exceed the critical size $\lambda/(2\pi)$ for both wavelengths,
leading to an increase in total opacity.
This enhanced opacity shifts the detected emission at the both wavelengths to the disk upper layers.
In this model,
strong irradiation extends the vertically isothermal region closer to the midplane.
At the same time, scattering is similarly effective at both wavelengths, resulting in comparable levels of scattering-induced attenuation.
As a result, the brightness temperatures converge,
and the spectral index approaches the blackbody value of $2$ (the bottom-right panel of Figure \ref{Multiwave}).

These results demonstrate that multi-wavelength observations, particularly at shorter wavelengths such as ALMA Band 7, provide a clear way to distinguish between the grain size limit and high-luminosity scenarios.
The predicted differences in both brightness temperature profiles and spectral indices at $0.88{\rm\:mm}$ will serve as a direct diagnostic of whether the grain size is limited or continues unrestricted under strong stellar irradiation.
Such observations will thus not only clarify which model better reflects the physical conditions of GGD27-MM1, but also advance our broader understanding of massive protostellar disks.

\subsection{Implications for Rocky Planetesimals Formation}\label{Planet}

While this study has focused on hot disks around massive protostars,
the derived constraints on the stickiness of silicate grains offer valuable insights into rocky planetesimal formation in protoplanetary disks around low-mass stars.

In all scenarios examined, the fragmentation velocity has been found to be consistently in the range of $12$--$17{\rm\:m\:s^{-1}}$.
This value is intermediate between the commonly adopted value of $\sim1{\rm\:m\:s^{-1}}$ and the extremely high value of $\sim50{\rm\:m\:s^{-1}}$ more recently predicted for aggregates made of $0.1~\mathrm{\mu m}$-sized dry silicates (see Section \ref{sec:intro}).
{Interestingly, this range of $v_{\rm frag} \approx 15{\rm\:m\:s^{-1}}$ is slightly larger but still comparable to the estimate based on observations of FU Orionis \citep{Liu2021}.}
Further analyses of additional hot disks around massive protostars and accretion-bursting stars will provide stronger constraints on the fragmentation velocity of silicate grains.

Although the estimated fragmentation velocity is higher than commonly adopted values, it is still lower than the typical maximum grain collisional velocity in protoplanetary disks around low-mass stars, which is $\approx20$--$50{\rm\:m\:s^{-1}}$ \citep[e.g.,][]{Birnstill2023}.
Additionally, the grain size limit scenario suggests that silicate grain growth might be limited to an even smaller size of $160{\rm\:\mu m}$.
Therefore, regardless of which scenario holds, our findings indicate that rocky planetesimals are unlikely to form solely through the collisional growth of silicate grains.
It is more likely that rocky planetesimal formation necessitates the gravitational collapse of dust overdensities formed by such as the streaming instability {\citep[e.g.,][]{Youdin+2005}}.

\subsection{Implications for Massive Star Formation}

Naturally, the methodology and findings of this study provide new insights into massive star formation, as our focus has been on hot disks surrounding massive protostars.
Over the past decade, theoretical and simulation studies have increasingly emphasized the critical role of accretion disks in the formation of massive stars.
These disks are essential for overcoming intense radiation feedback to enable the growth of massive stars \citep{Tanaka2017, Kuiper+18},
driving MHD outflows and jets \citep{Machida2020, Oliva2023},
and triggering accretion bursts and forming close companions through disk instability \citep{Meyer2024}.
On the observational front, ALMA has led to the discovery of accretion disks around massive protostars located several kpc away.
These observations have even revealed detailed structures within these disks \citep[e.g.,][]{Motogi2019, Johnston2020, Wright2022}
However, the impact of dust growth and fragmentation within disks on the formation of massive stars has remained largely unexplored.

\cite{Yamamuro2023} was the first to demonstrate how grain size evolution influences the physical structures of massive protostellar disks (see also Section \ref{sec:disk_property}).
In this study, we further show that grain growth also plays significant roles in shaping observational features at (sub-)mm wavelengths (Section \ref{sec:emission}),
which in turn affects the inferred protostellar parameters derived from continuum observations (Section \ref{sec:bestfit}).
When the maximum grain size exceeds the critical threshold $\lambda/(2\pi)$,
the increased opacity and scattering effects can reduce the disk brightness by $\sim20$--$30\:\%$.
This effect must be taken into account when estimating disk temperatures and accretion rates.
Additionally, spectral indices below 2 in two-wavelength observations near massive protostars have traditionally been interpreted as signatures of free-free or synchrotron emission, often used as indicators of photoionization and jet dynamics \citep{Sanna2018,Zhang2022}.
However, our results show that effective midplane heating and significant scattering effects can lower the spectral index to around 0.5, even when the emission originates solely from dust \citep[e.g.,][]{Liu2019, Sierra2020}.
Accurately extracting fundamental parameters of massive star formation, such as luminosity, accretion rates, and UV strength,
from observations plays a crucial role in constraining complex evolutionary models of massive protostars \citep{Hosokawa2009, Haemmerl2017A}.
The combination of disk modeling and radiative transfer calculations proposed in this study provides a powerful framework for comparing with other observations, offering valuable insights for future high-resolution studies of massive protostellar disks.

\subsection{Caveats}\label{c:4_1}

For GGD27-MM1, \cite{Anez+2020} identified a compact hot source ($T_{\rm B}\sim10^4{\rm K}$) within a few au of the disk, likely originating from ionized gas (see their Section 2.1).
To focus on the dust component, they excluded the contribution from this compact source and derived a corrected dust emission profile.
In this study, we used their corrected dust continuum data.
It is important to note, however, that the accuracy of this correction may affect the resulting brightness profile, particularly within the inner regions of $r \lesssim 50\:{\rm au}$.

We interpreted the polarization vectors pattern in ALMA Band 6 as originating from dust self-scattering.
However, the characteristic polarization pattern, i.e., the vectors aligned parallel to the disk's minor axis, is observed only in the {southwestern} region of the disk.
Therefore, we cannot rule out the possibility that the polarization may instead result from  the alignment of elongated grains due to magnetic fields or gas flows \citep[e.g.,][]{Anderson2015, Stephens2023}.
{If this is the case, estimates of the maximum grain size based on polarization would no longer be valid, rendering any inference of the fragmentation velocity highly uncertain.}
Unlike alignment-induced polarization, self-scattering polarization exhibits a strong wavelength dependence.
Multi-wavelength polarization observations are thus necessary to distinguish between these possible origins \citep{Ohashi2023}.

{ In this study, we adopt the plane-parallel slab approximation. However, since dust settling is not taken into account in our disk model, caution should be exercised when applying this approximation.}

We employed a dust model consisting of spherical and compact aggregates.
However, theoretical studies suggest that dust in protoplanetary disks may be fluffy porous aggregates \citep[e.g.,][]{Kataoka2013,Milchoulir2024}.
In addition, recent observations across infrared to radio wavelengths support the presence of porous dust structures in protoplanetary disks \citep{Tazaki2023, Zhang2023A, Ueda2024}.
The porosity of dust significantly impacts on planetesimal formation processes, including collisional growth \citep{Okuzumi+2012, Arakawa2023}.
Therefore, future constraints on the porosity of silicate dust in massive protostellar disks will be crucial.

We also adopted the ``astronomical silicate'' model for the dust composition, which is predominantly silicate.
However, (sub)milimeter observations have not detected any clear lines containing \ce{Si} compounds at the GGD27-MM1 disk\citep[e.g.,][] {Girart2017, Fernandez2023}.
Dust composition introduces significant uncertainties that strongly affect dust opacity and stickiness, both of which are crucial for grain growth and planetesimal formation.
Given these uncertainties, caution is warranted when interpreting the results of this study.
Future observations aimed at better constraining dust properties, including composition, will help improve the accuracy of model predictions.

\section{Summary} \label{sec:sum}

In this study, we compared our massive protostellar disk model with observed ALMA Band-6 observations of the hot disk around the massive protostar GGD27-MM1 to constrain the stickiness of silicate grains.
We self-consistently modeled the gas disk structure and dust coagulation, following the framework of \cite{Yamamuro2023},
where collisional fragmentation primarily regulates grain size.
We then performed radiative transfer calculations to examine how dust evolution influences on the millimeter-wave continuum emission from massive protostellar disks.
Our results showed that, when dust grains grow to sizes comparable to the observational wavelength,
strong scattering causes the disk brightness to fall below the blackbody radiation level corresponding to the disk surface temperature.

Using a model with a broad range of physical parameters, we aimed to reproduce the observed $1.14{\rm\:mm}$ continuum profile of the GGD27-MM1 disk.
The results indicated that, when attempting to match the dust distribution inferred from polarization,
the intensity in the inner region ($r \leq 100\:\mathrm{au}$) becomes approximately $20\:\%$ lower than the observed values,
due to increased opacity and effective scattering by large grains.
To resolve this discrepancy, we proposed two possible scenarios:
(1) the ``grain size limit scenario,'' where dust growth is limited to $\sim100\:\mathrm{\mu m}$ due to bouncing,
and the ``high-luminosity scenario,'' which accounts for uncertainties in the stellar luminosity.
These two scenarios can be distinguished through follow-up observations at ALMA Band 7 ($0.88{\rm\:mm}$).

Based on the inferred properties of silicate grains in the massive protostellar disk,
we discussed the implications for rocky planetesimal formation in protoplanetary disks around low-mass stars.
In both scenarios, the fragmentation velocity of silicate dust was found to be $12$--$17{\rm\:m\:s^{-1}}$, which is larger than the commonly used value of $1{\rm\:m\:s^{-1}}$.
However, this value is still lower than the typical maximum collisional velocity of $20$--$50{\rm\:m\:s^{-1}}$ in protoplanetary disks,
implying that rocky planetesimal formation through collisional growth alone is unlikely.
This difficulty is particularly pronounced in the case of the grain size limit scenario,
where grain growth is restricted to around $160{\rm\:\mu m}$.

\section*{}
{This work refers to observational results from \citet{Anez+2020}, which made use of ALMA data (ADS/JAO.ALMA\#2015.1.00480.S).
ALMA is a partnership of ESO (representing its member states), NSF (USA), and NINS (Japan), together with NRC (Canada), NSTC and ASIAA (Taiwan), and KASI (Republic of Korea), in cooperation with the Republic of Chile. The Joint ALMA Observatory is operated by ESO, AUI/NRAO, and NAOJ.}
The authors are grateful to Satoshi Ohashi, Ryoki Matsukoba, and Kazuhito Motogi for useful discussions.
This work was supported by JSPS KAKENHI Grant Numbers JP20H00182, JP23H00143, JP23K25923, JP21H01145, and JP25K07365.
This work was supported by Yoshinori Ohsumi Fund (Yoshinori Ohsumi Award for Fundamental Research), the Hayakawa Satio Fund awarded by the Astronomical Society of Japan, and the Hiki Foundation by Institute of Science Tokyo.
{We thank the anonymous referee for comments, which were useful to improve the original manuscript.}

\appendix

\section{All models of the Fiducial Case}\label{app:fid}
\begin{figure}[b]
\begin{center}
\includegraphics[width=\hsize]{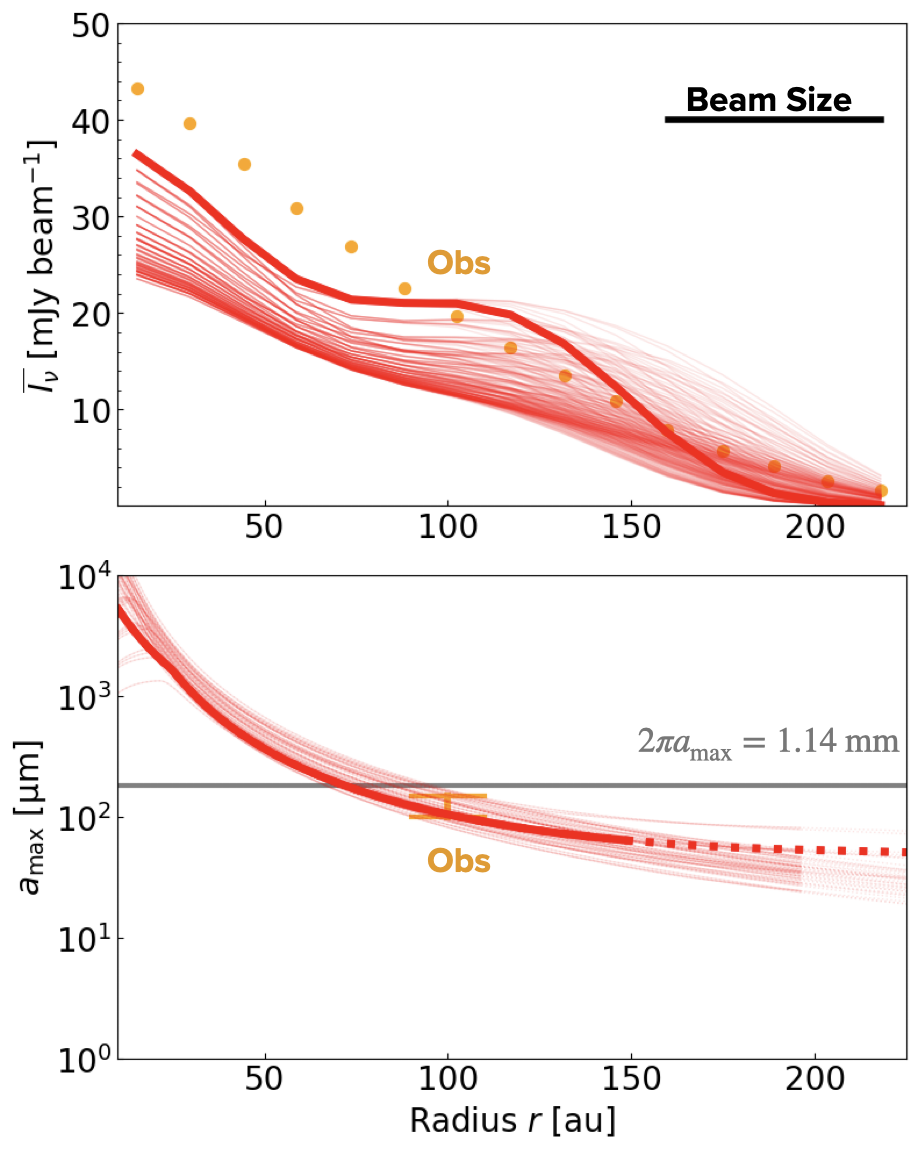}\
\caption{The same as Figure \ref{Best_fit}, but for the all models in Section \ref{sec:Bestfit}. }
\label{Hazure}
\end{center}
\end{figure}

The fit model failed to reproduce the luminosity of the inner disk in the observations in Section \ref{sec:Bestfit}.
In this section, we show the all the models investigated in Section \ref{sec:Bestfit} in Figure \ref{Hazure}.  
In all models, the brightness $I_\nu$ within the inner region of the disk $\lesssim 100 \:\mathrm{au}$ is $\gtrsim20\:\%$ dimmer compared to the observations.

\section{Models for the GGD27-MM1 disk} \label{sec:appA}
\begin{figure*}
\includegraphics[width=\hsize]{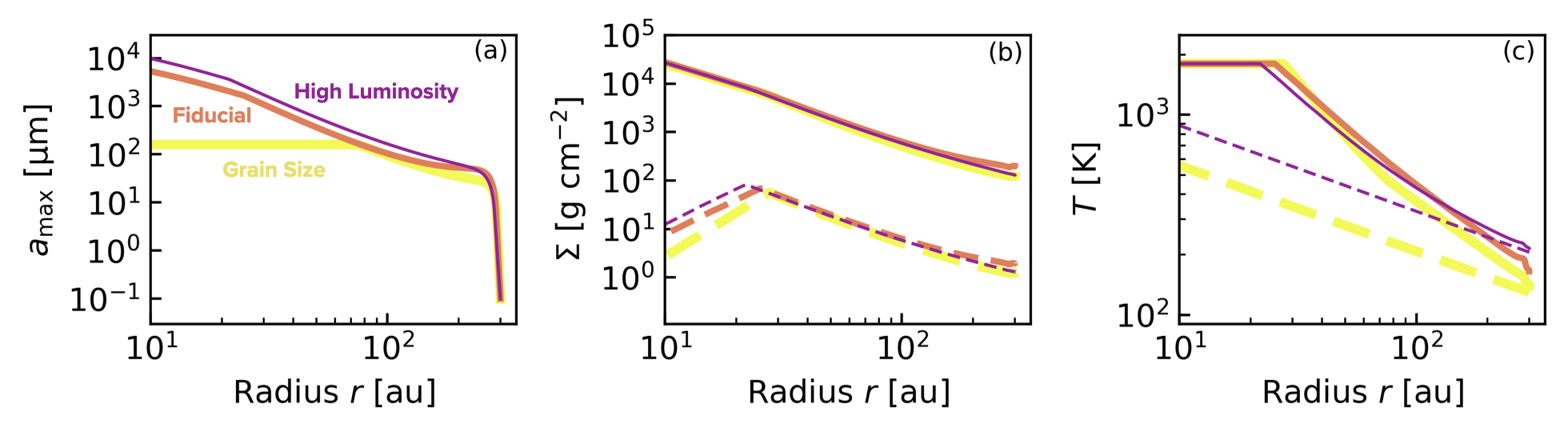}\
\caption{
Radial profiles of physical quantities in the fiducial model (orange lines), the grain size limit case (yellow lines) and the high luminosity case (purple lines):
(a) maximum grain radius $a_{\mathrm {max}}$;
(b) surface densities of gas $\Sigma_\mathrm{gas}$ (solid) and dust $\Sigma_{\mathrm {dust}}$ (dashed);
(c) midplane temperature $T_\mathrm{mid}$ and the surface temperature $T_{\rm irr}$ (dashed).
}
\label{ap_disk}
\end{figure*}

In Sections \ref{sec:Bestfit} and \ref{43}, we presented three fitted models.  
We show the physical properties of each disk model in Figure \ref{ap_disk}.
The basic behaviors of all physical quantities are the same as the reference model in Section \ref{sec:disk_property}.
The grain size limit case and high luminosity case showed similar results in the $1.14\:\mathrm{mm}$ continuum emission.
However, as discussed in Section \ref{Multiwave}, differences in dust growth limitations and stellar luminosity lead to distinct trends in the radial distribution of grain size and disk temperature.

\section{The Dust Opacity as a function of Maximum Grain Size} \label{ap:dust-op}
Finally, we introduce the relationship between dust growth and dust opacity.
As demonstrated throughout this study, dust growth leads to variations in dust opacity at sub-milimetere range.
In this section, we provide an example of the impact of opacity changes due to dust growth not only in the ALMA wavelength range.
To present the results concisely, we show the opacities for maximum grain radii of $1\:\mathrm{\mu m}$, $100\:\mathrm{\mu m}$, and $1\:\mathrm{cm}$ {where the dust model is the same as that described in Section\:\ref{model:op}}.

\begin{figure*}
\includegraphics[width=\hsize]{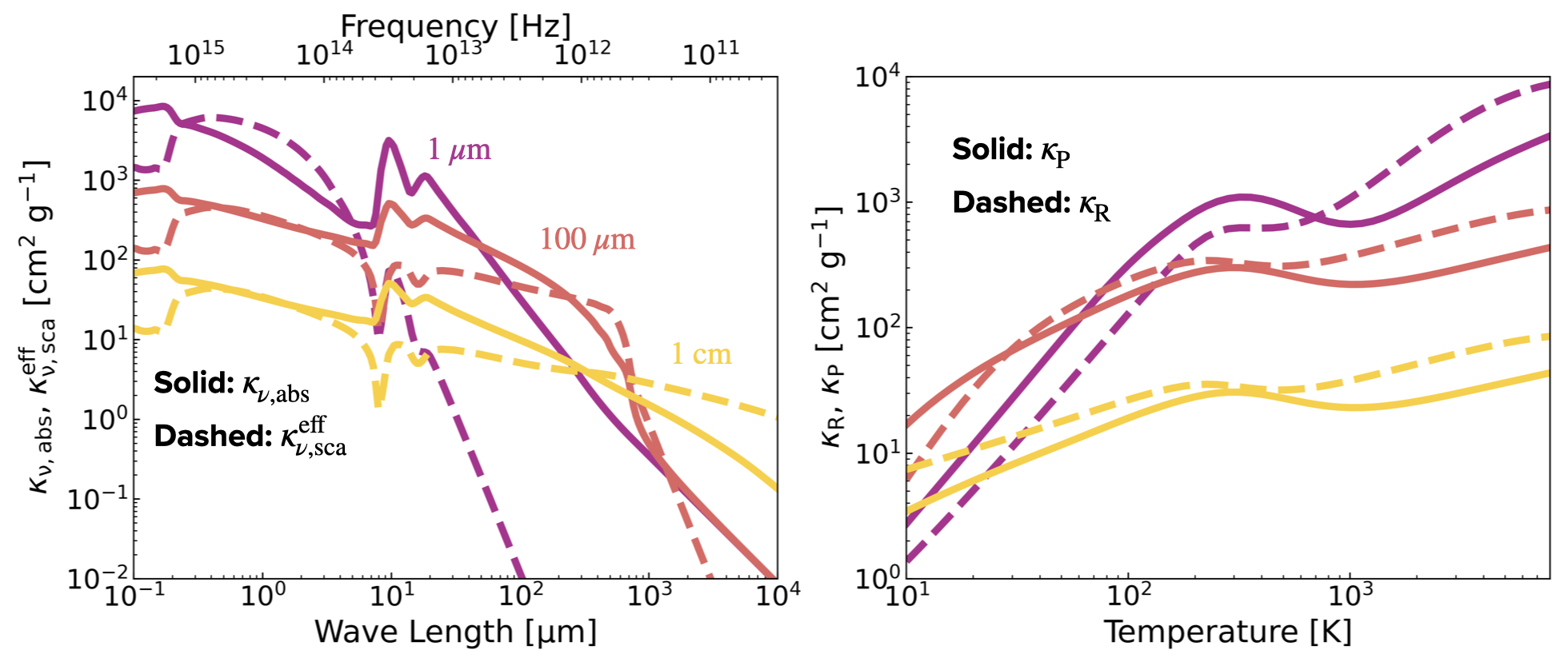}\
\caption{The opacity profiles of the maximum grain radius of $1\:\mathrm{\mu m}, 100\:\mathrm{\mu m},$ and $1\:\mathrm{cm}$ (purple, orange, and yellow lines), respectively.
Left: the absorption opacity $\kappa_\mathrm{\nu, abs}$ (solid lines) and effective scattering opacity $\kappa^\mathrm{eff}_\mathrm{\nu,\:sca}$(dashed) per dust mass as the function of wavelength and frequency.
Right: the Planck mean opacity $\kappa_\mathrm{P}$ (solid lines) and Rosseland mean opacity $\kappa_\mathrm{R}$ (dashed lines) per {dust} mass as the function of temperature. 
{Note that the grain size distribution is assumed to follow a power law with an exponent of $-3.5$, with a minimum grain radius of $0.1\:\mathrm{\mu m}$. These settings are consistent with those used for the dust opacity described in Section \ref{model:op} for details.}
}
\label{op_wide}
\end{figure*}

The light panel of Figure\:\ref{op_wide} shows the absorption and scattering opacities at $100\:\mathrm{nm}$--$1\:\mathrm{cm}$ wavelength of each maximum grain radius.
The opacities depend on the maximum grain radius.
For wavelengths shorter than $2\pi a_\mathrm{max}$, the opacity increases as the dust grain radius decreases.
However, for wavelengths longer than $2\pi a_\mathrm{max}$, the opacity decreases with smaller dust grain radii and eventually converges to a similar value.
Furthermore, as discussed in Section \ref{model:op}, scattering becomes significant at wavelengths around $\lambda \simeq 2\pi a_\mathrm{max}$ across all wavelength ranges.
However, a distinct feature appears around $\lambda\sim 10\:\mathrm{\mu m}$, where opacity increases sharply.  
This is due to the use of astronomical silicate, which causes a characteristic silicate absorption band at $10\:\mathrm{\mu m}$.  

In addition,to demonstrate the dust growth effect on dust opacity, we show the Rosseland and Planck mean opacities as a function of temperature for the maximum grain size at the right panel of Figure\:\ref{op_wide}.
The Planck mean opacity is evaluated as
\begin{equation}\label{eq:kappa_planck}
{\kappa_\mathrm{P}(a_\mathrm{max},T)}=
\cfrac{\displaystyle \int^\infty_0 \kappa_{\nu,\mathrm{abs}} B_\nu(T) d \nu} {\displaystyle \int^\infty_0  B_\nu(T) d \nu}.
\end{equation}
The Rosseland mean and Planck mean opacities tend to increase with temperature.
Additionally, while smaller grains generally result in higher opacity, this relationship reverses at certain wavelengths.
This reversal is related to the wavelength at which the Planck function $B_\nu$ reaches its maximum for a given temperature $T$, corresponding to the wavelength that carries the most energy, $\lambda_\mathrm{max} = 10\:(300\:\mathrm{K}/T)\:\mathrm{\mu m}$, as well as the relationship between the maximum grain radius.
As a result, in the temperature range of $10$--$50\:\mathrm{K}$, where $\lambda_\mathrm{max} \sim 60$--$300\:\mathrm{\mu m}$, dust with a maximum grain radius of $a_\mathrm{max}=100\:\mathrm{\mu m}$ exhibits higher opacity than dust with $a_\mathrm{max}=1\:\mathrm{\mu m}$.
In the temperature range of $300$--$1000\:\mathrm{K}$, a wave-like structure in the opacity is observed for all maximum grain radius.
Considering the wavelength corresponding to $\lambda_\mathrm{max}$ in this temperature range, this structure can be attributed to $10\:\mathrm{\mu m}$ feature of silicate.

These result indicate that dust growth plays an important role in both the observation of massive protostellar disks at each wavelength and in several radiative processes, such as disk heating, radiation pressure from the central star, and the photoionization rate.

\bibliography{ms}{}
\bibliographystyle{aasjournal}

\end{document}